\documentclass[twocolumn]{aastex63}
 \usepackage{CJKutf8}
\usepackage{hyperref}
\usepackage{pbox}
\usepackage{amsmath}
\usepackage{makecell}
\graphicspath{{./}{figures/}}
\usepackage{booktabs}
\newcommand{\cntext}[1]{\begin{CJK}{UTF8}{gbsn}#1\end{CJK}\kern-1ex}
\newcommand{\uat}[2]{\href{http://astrothesaurus.org/uat/#2}{#1 (#2)}}
\shorttitle{Energetic Electrons Accelerated in a Solar Flare Magnetic Bottle}
\shortauthors{Chen et al.}

\begin{document}

\title{Energetic Electrons Accelerated and Trapped in a Magnetic Bottle above a Solar Flare Arcade} 

\author[0000-0002-0660-3350]{Bin Chen (\cntext{陈彬})}
\affiliation{Center for Solar-Terrestrial Research, New Jersey Institute of Technology, 323 Martin Luther King Blvd., Newark, NJ 07102-1982, USA}

\author[0000-0003-1034-5857]{Xiangliang Kong}
\affiliation{Institute of Space Sciences, Shandong University, Weihai, Shandong 264209, China}

\author[0000-0003-2872-2614]{Sijie Yu}
\affiliation{Center for Solar-Terrestrial Research, New Jersey Institute of Technology, 323 Martin Luther King Blvd, Newark, NJ 07102-1982, USA}

\author[0000-0002-9258-4490]{Chengcai Shen}
\affiliation{Harvard-Smithsonian Center for Astrophysics, Cambridge, MA 02138, USA}

\author[0000-0001-5278-8029]{Xiaocan Li}
\affiliation{Department of Physics and Astronomy, Dartmouth College, Hanover, NH 03755, USA}

\author[0000-0003-4315-3755]{Fan Guo}
\affiliation{Los Alamos National Laboratory, Los Alamos, NM 87545, USA}

\author[0000-0001-8941-2017]{Yixian Zhang}
\affiliation{School of Physics \& Astronomy, University of Minnesota Twin Cities, Minneapolis, MN 55455, USA}

\author[0000-0001-7092-2703]{Lindsay Glesener}
\affiliation{School of Physics \& Astronomy, University of Minnesota Twin Cities, Minneapolis, MN 55455, USA}

\author[0000-0002-2002-9180]{S\"am Krucker}
\affiliation{Space Sciences Laboratory, University of California, 7 Gauss Way, Berkeley, CA 94720, USA}
\affiliation{University of Applied Sciences and Arts Northwestern Switzerland, 5210 Windisch, Switzerland}

\begin{abstract}
Where and how flares efficiently accelerate charged particles remains an unresolved question. Recent studies revealed that a ``magnetic bottle'' structure, which forms near the bottom of a large-scale reconnection current sheet above the flare arcade, is an excellent candidate for confining and accelerating charged particles. 
However, further understanding its role requires linking the various observational signatures to the underlying coupled plasma and particle processes. 
Here we present the first study combining multiwavelength observations with data-informed macroscopic magnetohydrodynamics and particle modeling in a realistic eruptive flare geometry. The presence of an above-the-loop-top magnetic bottle structure is strongly supported by the observations, which feature not only a local minimum of magnetic field strength but also abruptly slowing down plasma downflows. It also coincides with a compact hard X-ray source and an extended microwave source that bestrides above the flare arcade. Spatially resolved spectral analysis suggests that nonthermal electrons are highly concentrated in this region.
Our model returns synthetic emission signatures that are well matched to the observations. The results suggest that the energetic electrons are strongly trapped in the magnetic bottle region due to turbulence, with only a small fraction managing to escape. The electrons are primarily accelerated by plasma compression and facilitated by a fast-mode termination shock via the Fermi mechanism. Our results provide concrete support for the magnetic bottle as the primary electron acceleration site in eruptive solar flares. They also offer new insights into understanding the previously reported small population of flare-accelerated electrons entering interplanetary space.
\end{abstract}


\section{Introduction}\label{sec:intro}
The discovery of hard X-ray (HXR) sources located above the bright solar flare arcade (after Masuda et al. \citealt{1994Natur.371..495M}) has convincingly placed the primary flare energy release and particle acceleration site to the coronal volume. It coincides with the location where a large-scale current sheet is present to drive the flare energy release via magnetic reconnection---a process in which magnetic field lines break and reconnect to unleash the previously stored magnetic energy. Owing to its strong electric field that can reach thousands of volts per meter, this reconnection current sheet has often been suggested as the main driver for particle acceleration \citep{1988ApJ...330L.131M,1996ApJ...462..997L, 2000A&A...360..715K, Drake2006,2011ApJ...737...24B,2018ApJ...866....4L}. 

\begin{figure*}[!ht]
\begin{center}
\includegraphics[width=0.7\textwidth]{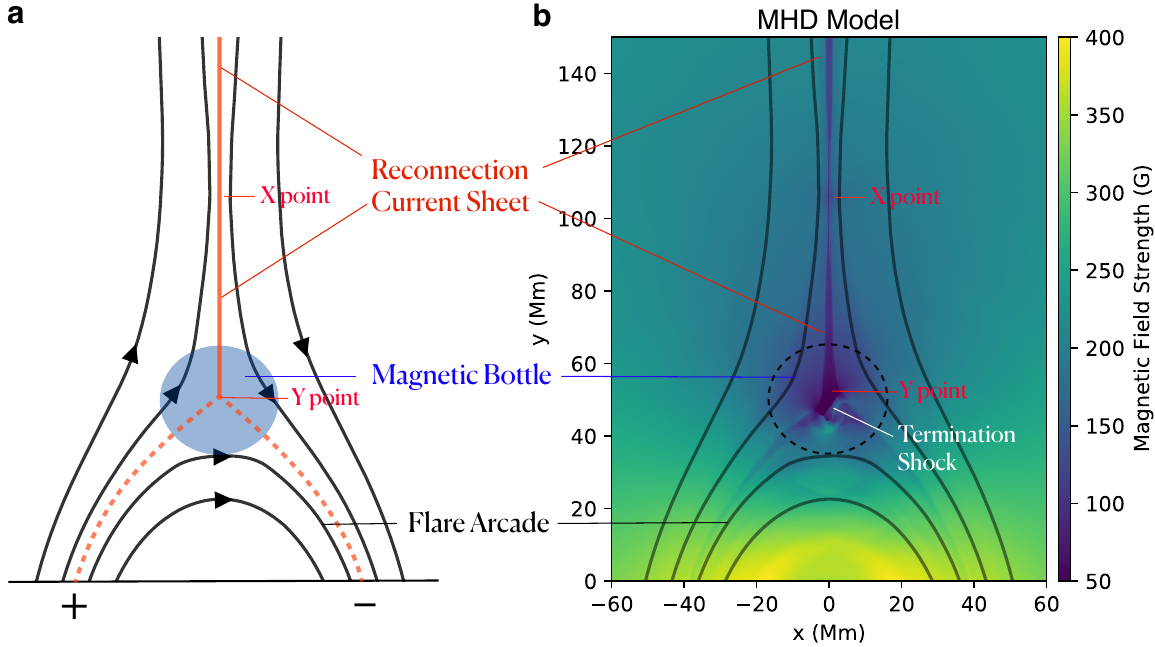}
\end{center}
\caption{Formation of a magnetic bottle structure above a solar flare arcade. (a) The presence of a large-scale reconnection current sheet above the flare arcade leads to a local divergence of the magnetic field lines in the above-the-loop-top region, forming a magnetic bottle (shaded blue ellipse). The vertical red line is the reconnection current sheet, and the dashed red curve denotes the quasi-separatrix layer. (b) The total magnetic field strength distribution in a 2.5D resistive MHD model shows a generally lower magnetic field strength in the magnetic bottle region. Also marked are the X point where the field lines break and reconnect, the Y point where the current sheet meets the quasi-separatrix layer, and a fast-mode termination shock formed by reconnection outflows impinging upon the flare arcade.}\label{fig:b_bottle}
\end{figure*}

However, since energetic particles are extremely mobile in the solar corona,  the acceleration of a large number of particles to nonthermal energies requires efficient bulk acceleration, strong trapping, or both. Where and how such bulk acceleration and trapping occur remains an unresolved problem. Moreover, studies that combine \textit{in situ} spacecraft measurements and remote-sensing observations have concluded consistently that only 0.1--1\% of the flare-accelerated energetic electrons manage to escape to interplanetary space \citep{Lin1974,Krucker2007,Dresing2021,Wang2021,WangM2023}. Such a profound departure from equipartition between the upward-escaped and downward-precipitated/trapped energetic electrons at the flare site has posed a significant challenge in understanding the particle acceleration and transport processes.

Previous results based on HXR analysis of the electron time-of-flight distances \citep{Aschwanden1996a, Aschwanden1996b} and observations of above-the-loop-top (hereafter ALT) sources with a high nonthermal electron density \cite[e.g.,][]{Krucker2010, Ishikawa2011,Krucker2014,Fleishman2022} have suggested that the primary acceleration in large eruptive flares may be located in the cusp region just above the flare arcade but not necessarily in the upper portion of the current sheet, including the primary X point(s). 
Recently,  by combining microwave imaging spectroscopy observations with magnetohydrodynamics (MHD) modeling, \citet{Chen2020NatAs} found that the region near the bottom of the current sheet (also known as the ``Y point'' owing to the bifurcation of the current layer; see Figure~\ref{fig:b_bottle}(a)) coincides with a local depression of magnetic field strength. This peculiar structure, referred to as a ``magnetic bottle,'' is a natural consequence of energy release driven by magnetic reconnection in a large-scale, vertical current sheet above the flare arcade. In the schematic picture shown in Figure~\ref{fig:b_bottle}(a), the antiparallel magnetic field lines encompassing the current sheet diverge toward the top of the flare arcade. A similar field line divergence is also evident as one follows the footpoints of the flare arcade upward toward the loop top. In turn, the conservation of magnetic flux demands a general reduction of the magnetic field strength in the ALT region. Such a physical picture is supported by analytical models of flare reconnections \citep{Fletcher1998trap,2000JGR...105.2375L} and numerical simulations that solve the MHD equations. Figure~\ref{fig:b_bottle}(b) shows an example frame from our MHD simulations (after \citealt{Shen2018}), which displays a reduced magnetic field strength at the same region.

\begin{figure*}[!ht]
\begin{center}
\includegraphics[width=0.62\textwidth]{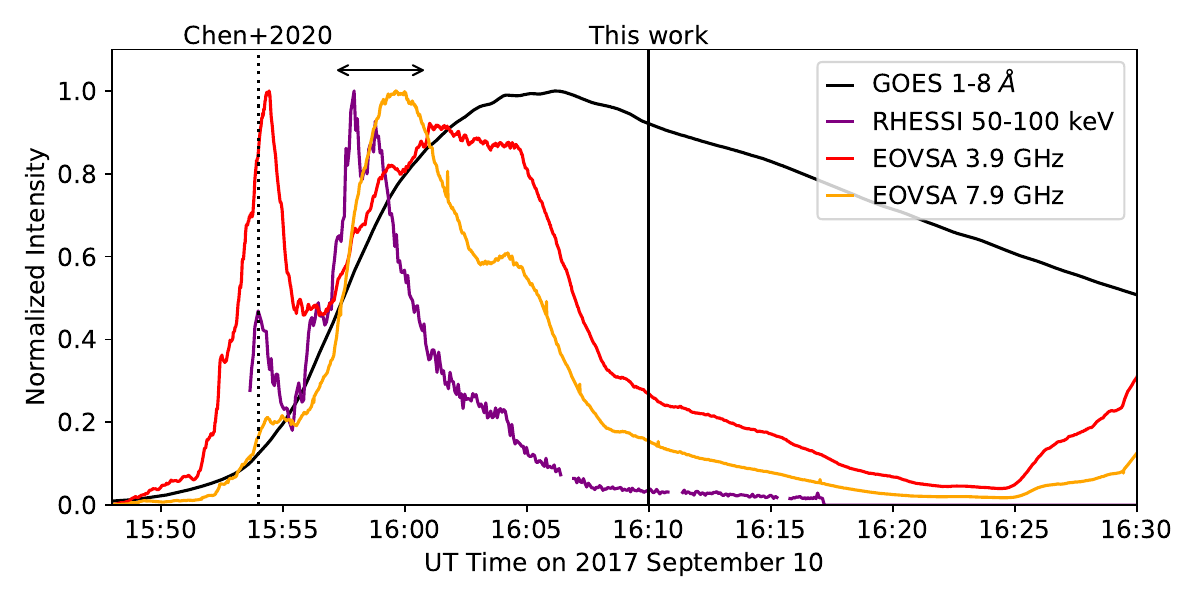}
\end{center}
\vskip -15pt
\caption{Microwave and X-ray light curves of the X8.2-class eruptive solar flare on 2017 September 10. The solid vertical line indicates the time of interest of this work during the gradual phase of the flare (around 16:10 UT). The dotted vertical line marks the time studied by \citet{Chen2020NatAs} during the early impulsive phase of the flare when the eruption was initiated, and the double-sided arrow denotes the main impulsive phase of the flare around the primary HXR and microwave peak (see \citealt{Gary2018eovsa} for more details on the flare evolution).}\label{fig:lc}
\end{figure*}

This magnetic bottle structure coincides with the location of the ALT HXR source and shows a strong concentration of microwave-emitting nonthermal electrons \citep{Chen2020NatAs}, implicating a key role it may be playing in the particle acceleration processes. This is also where the fast reconnection outflows---which carry the bulk of the released energy---collide head-on against the flare arcade to create a plethora of energetic phenomena such as collapsing magnetic traps \citep{1997ApJ...485..859S, Karlicky2004}, fast-mode termination shocks \citep{Forbes1986, Tsuneta1998, Aurass2002, Aurass2004, Mann2009, Guo2012, Chen2015, Chen2019, Takasao2015, Polito2018, Shen2018, Shen2022, Ye2020, Luo2021, French2024}, slow-mode or gas dynamic shocks \citep{Reeves2007, Longcope2011looptop, Longcope2016, Longcope2022}, and turbulence and oscillations \citep{Takasao2016, Kontar2017, Reeves2020, Ye2020, Shen2022, Shen2023, Ruan2023, Shibata2023, WangY2023}, serving as an ideal environment to heat flare plasma and accelerate charged particles.

However, understanding how the local concentration occurs and, in turn, the underlying particle acceleration and transport mechanisms requires a concerted observational-modeling approach that links the various emission features to the highly coupled plasma dynamics and particle processes, which has been heretofore elusive. Here, we use a novel macroscopic MHD and particle model (after \citealt{Kong2019}) to produce not only a distribution of thermal plasma in a realistic flare geometry but also a spatially, spectrally, and temporally resolved distribution of energetic electrons throughout the flare region. The model makes it possible, for the first time, to compare the model outputs with multiwavelength imaging spectroscopy observations that trace both the heated plasma and nonthermal electrons. The paper is organized as follows. Section~\ref{sec:obs} discusses the multiwavelength observations and analysis. Section~\ref{sec:model} describes the MHD, particle, and emission modeling setup and results. Section~\ref{sec:discussion} puts both the observational and modeling results into a coherent physical context and discusses their implications. For the sake of readability, extensive technical details for X-ray and microwave data analysis and numerical modeling are included in the Appendices.

\begin{figure*}[!ht]
\begin{center}
\includegraphics[width=0.7\textwidth]{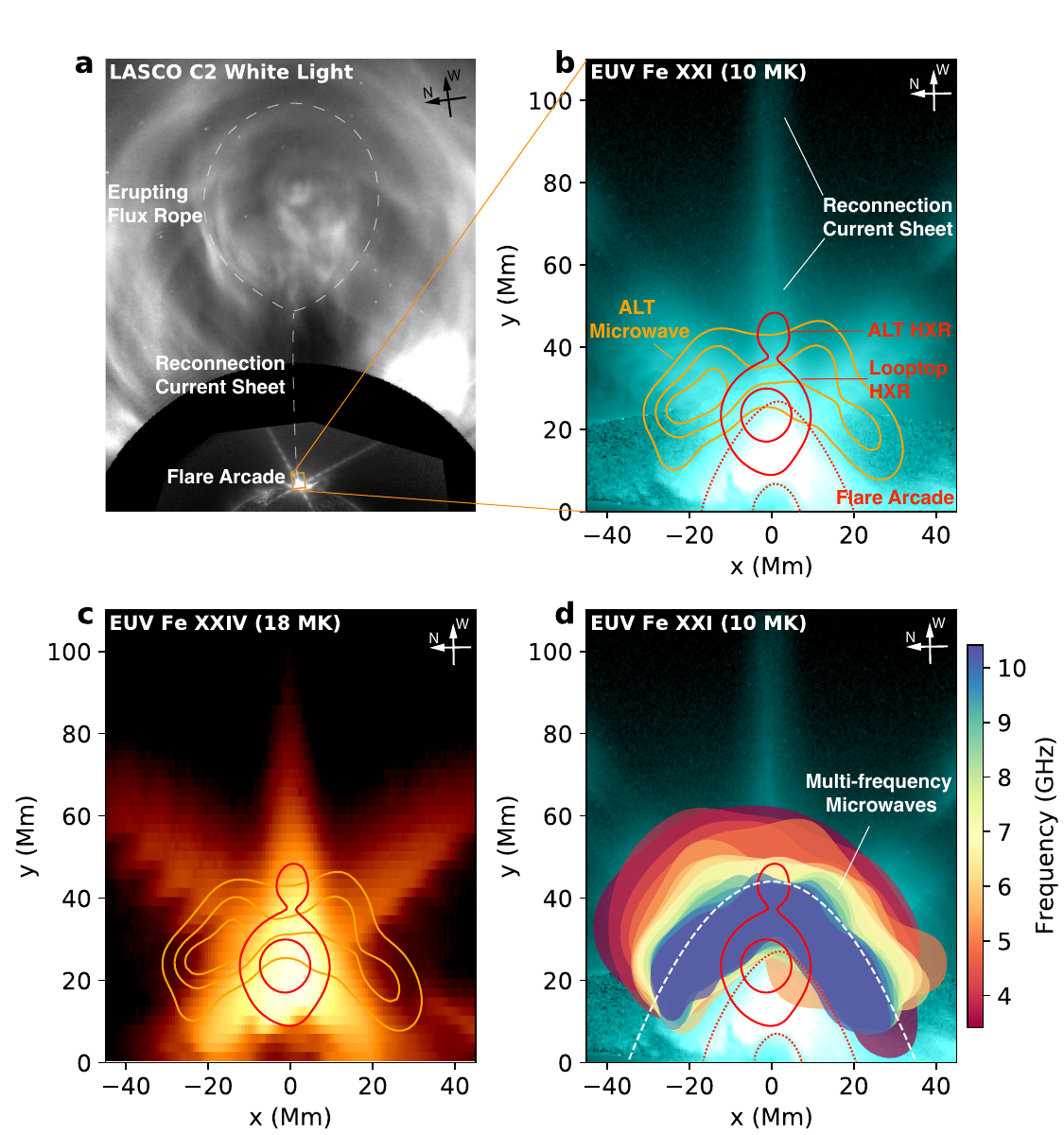}
\end{center}
\vskip -15pt
\caption{Multiwavelength observations of the flare during its gradual phase at 16:10 UT. (a) A CME observed in white light by the SOHO/LASCO coronagraph along with a trailing long plasma sheet. The inset is the 131 \AA\ EUV channel image made by the Solar Ultraviolet Imager (SUVI) on board the GOES-R satellite. (b) Detailed view of the plasma sheet and postflare arcade region. A small box denotes the field of view in (a). The background is the SDO/AIA 131 \AA\ EUV channel image, which samples the hot 10 MK plasma through the \ion{Fe}{21} line, shown with an inverse color scale (i.e., darker color is brighter). Orange contours are the 7.9 GHz microwave source (15\%, 50\%, and 90\% of the maximum). The RHESSI 25--60 keV HXR image is shown as red contours (10\% and 50\% of the maximum). (c) Same as (b), but the background is the \ion{Fe}{24} line EUV image observed by Hinode/EIS, which is sensitive to 18 MK plasma. (d) Similar to (b) but with filled contours denoting multifrequency EOVSA microwave images from 3.4 to 12.9 GHz. }\label{fig:overview}
\end{figure*}

\section{Observations}\label{sec:obs}
The observations were obtained during the well-observed X8.2-class eruptive solar flare on 2017 September 10, recorded by ground- and space-based instruments at multiple wavelengths. 
This event, thanks to its favorable viewing perspective, has a geometry that matches very well the standard model of eruptive solar flares (Figure~\ref{fig:overview}). We refer interested readers to \citet{Chen2020fluxrope} and references therein for a more detailed discussion on its three-dimensional (3D) configuration based on multiwavelength, multiperspective observations. Briefly, the event is induced by an erupting magnetic flux rope that drives a fast white-light coronal mass ejection (CME). Immediately trailing the CME core, a long, linear feature, seen in both white light and extreme ultraviolet (EUV), extends down to the top of the bright flare arcade, interpreted as a large-scale reconnection current sheet viewed edge-on \citep{Warren2018cs,Yan2018,Chen2020NatAs}. The flare arcade anchored at the solar surface displays a cusp shape as observed by the Atmospheric Imaging Assembly (AIA; \citealt{2012SoPh..275...17L}) on board the Solar Dynamics Observatory (SDO) and the EUV Imaging Spectrometer (EIS; \citealt{Culhane2007}) on board Hinode (see Figures~\ref{fig:overview}(b) and (c), respectively). The cusp-shaped flare loops are the signature of highly bent magnetic field lines resulting from ongoing magnetic reconnection in the current sheet, producing fast sunward reconnection outflows \citep[e.g.,][]{Longcope2018downflows,Hayes2019,Yu2020outflows}. 

In order to carry out the comparison with our combined MHD, particle, and emission model (see Section~\ref{sec:model}), we select a period around 16:10 UT during the gradual phase of the eruptive solar flare when the eruption has already propagated to a large distance (i.e., 12 minutes after the main microwave and HXR peak at around 15:58 UT; Figure~\ref{fig:lc}). As shown in Figure~\ref{fig:overview}(a), the core of the white-light CME is located at $\sim$4 $R_{\odot}$, followed by a long trailing plasma sheet. In comparison, the size of the flare arcade itself is only $\sim$20 Mm, or $\sim$1\% of the length of the sheet. Therefore, the plasma processes near the solar surface can be well approximated by a system driven by an infinitely long current sheet as depicted in Figure~\ref{fig:b_bottle}, which allows us to perform detailed data--model comparison in a realistic flare geometry. 

We note that the flare event occurred on the west solar limb, and the eruption generally proceeded along the east--west direction. For the sake of easier data--model comparison, throughout this paper, we have rotated the observed images by 90$^{\circ}$ counter-clockwise and assigned $x=950''$ and $y=-140''$ in the original helioprojective Cartesian coordinates as the origin of our new coordinate system. The length units are in megameters, with $1''\approx 0.73$ Mm at a distance of 1 AU.

\subsection{X-ray and Microwave Imaging and Spectroscopy}

\begin{figure}[!ht]
\begin{center}
\includegraphics[width=0.45\textwidth]{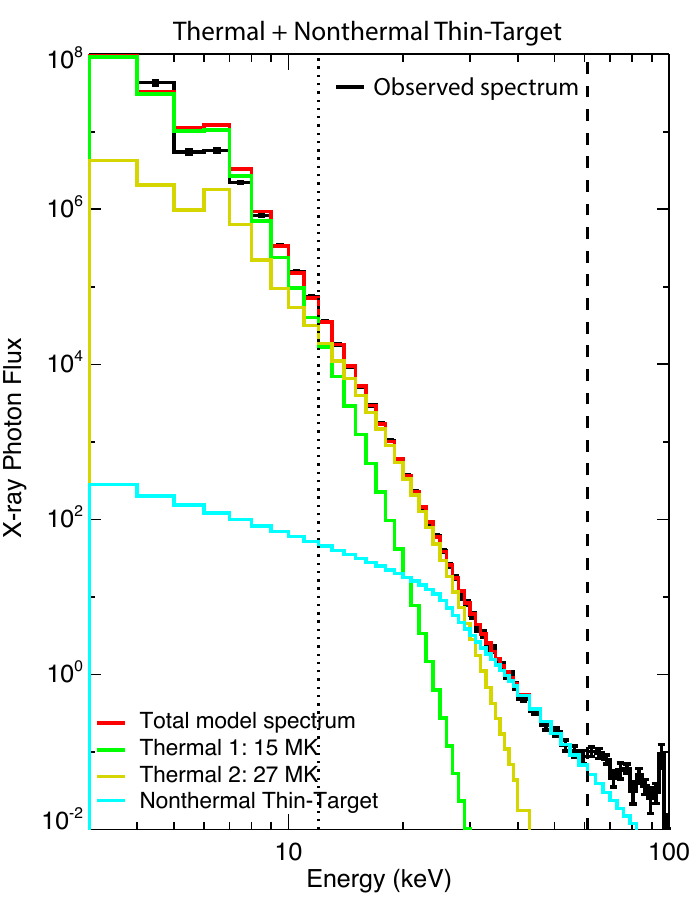}
\end{center}
\caption{RHESSI X-ray spectral analysis results for the time of interest at 16:10 UT. (a) The black curve represents the observed X-ray photon spectrum averaged over 16:10:00 UT and 16:11:08 UT. The red curve is the best-fit spectrum that includes two isothermal components and one nonthermal component arising from thin-target bremsstrahlung from a power-law electron distribution. The two thermal components, shown as the green and yellow curves,
have temperatures of 15 MK and 27 MK and volume emission measures of $1.6\times 10^{51}\,\mathrm{cm}^{-3}$ and $4.1\times 10^{49}\,\mathrm{cm}^{-3}$, respectively. The nonthermal thin-target component has a normalization factor of $1.6\times 10^{54}\,\mathrm{cm}^{-2}\,\mathrm{s}^{-1}$ with a power-law index of $\delta^{\rm thin}=4.7$.}\label{fig:xray_spec}
\end{figure}

In the X-ray images obtained by the Reuven Ramaty High Energy Solar Spectroscopic Imager (RHESSI; \citealt{Lin2002rhessi}), this period features an ALT HXR source at 25--60 keV (Figure~\ref{fig:overview}(b); see also Figure~\ref{fig:xray_img} in Appendix for images at additional energy bands). At the top of the EUV flare arcade, a brighter X-ray source is present (referred to as the ``loop-top'' source), which is dominated by thermal bremsstrahlung from hot ($>$15 MK) plasma with a high density ($\sim\!10^{12}\,\mathrm{cm}^{-3}$). In the total-flux X-ray spectrum, at above $\sim$30 keV, a nonthermal component dominates the spectrum with a power-law shape (Figure~\ref{fig:xray_spec}). Spectral analysis suggests that if this component falls into the thin-target bremsstrahlung regime, it corresponds to a total nonthermal electron density of $n_{e}^{>50}\approx 4\times10^{6}$~cm$^{-3}$ above 50 keV (see detailed discussions in Appendix~A). 

\begin{figure*}[!ht]
\begin{center}
\includegraphics[width=0.48\textwidth]{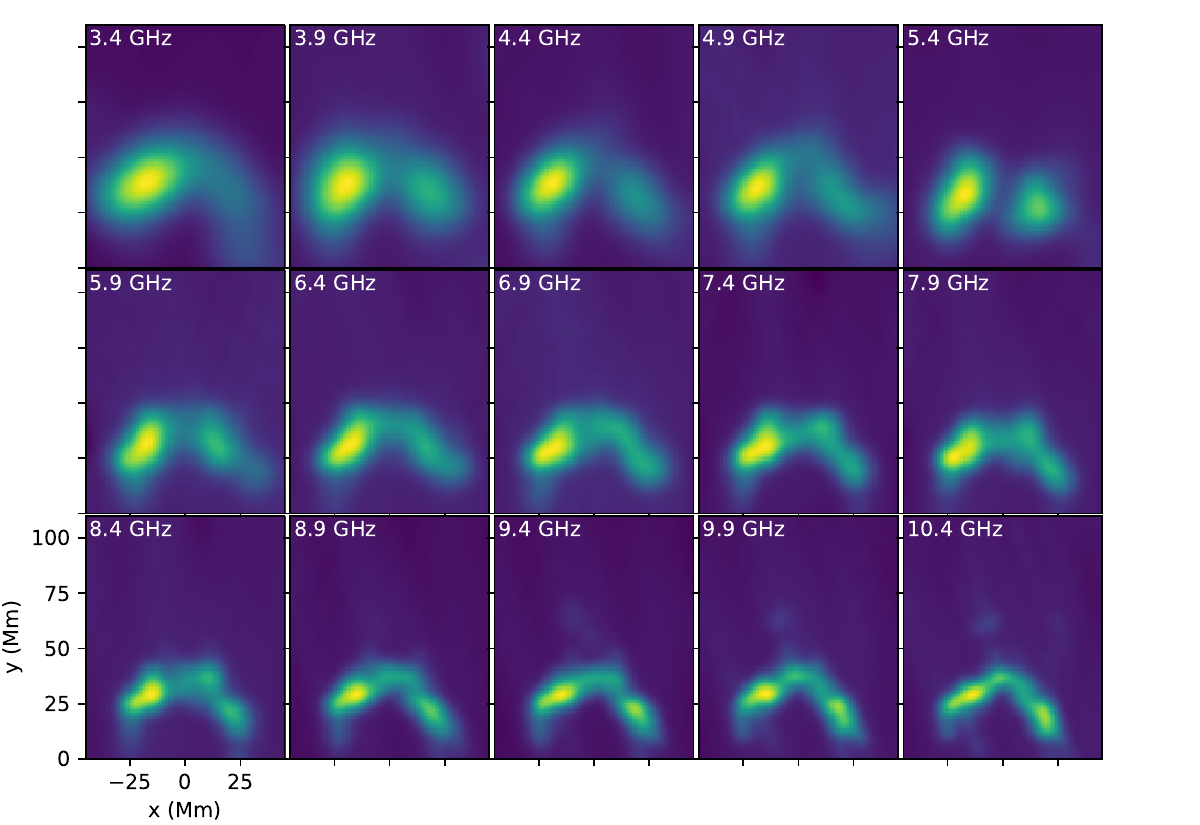}
\includegraphics[width=0.48\textwidth]{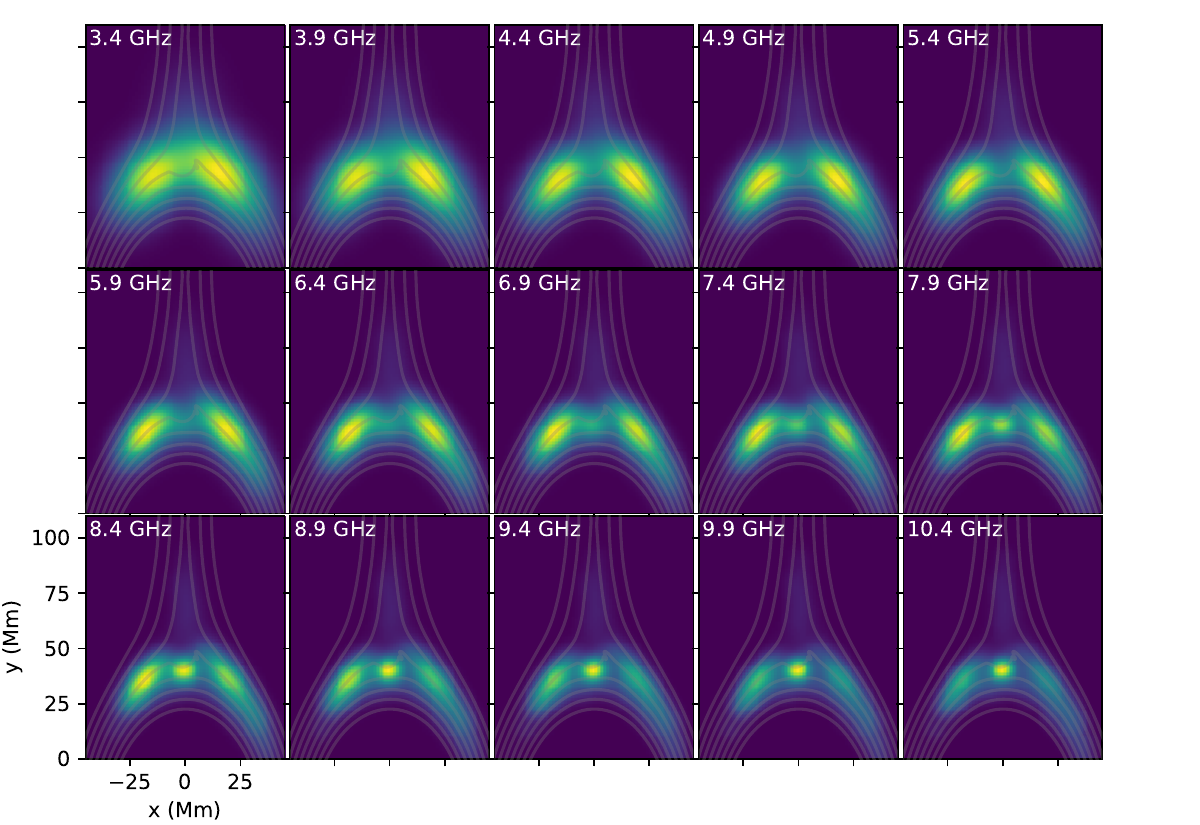}
\end{center}
\caption{Observed and modeled multifrequency microwave images from 3.4 to 10.4 GHz. Left panels: microwave images observed by EOVSA on 2017 September 10 at 16:10:36 UT. Right panels: synthetic EOVSA microwave images generated from the combined MHD and particle model. Overlaid gray curves are magnetic field lines derived from the MHD model. }\label{fig:microwave_images}
\end{figure*}

\begin{figure*}[!ht]
\begin{center}
\includegraphics[width=0.9\textwidth]{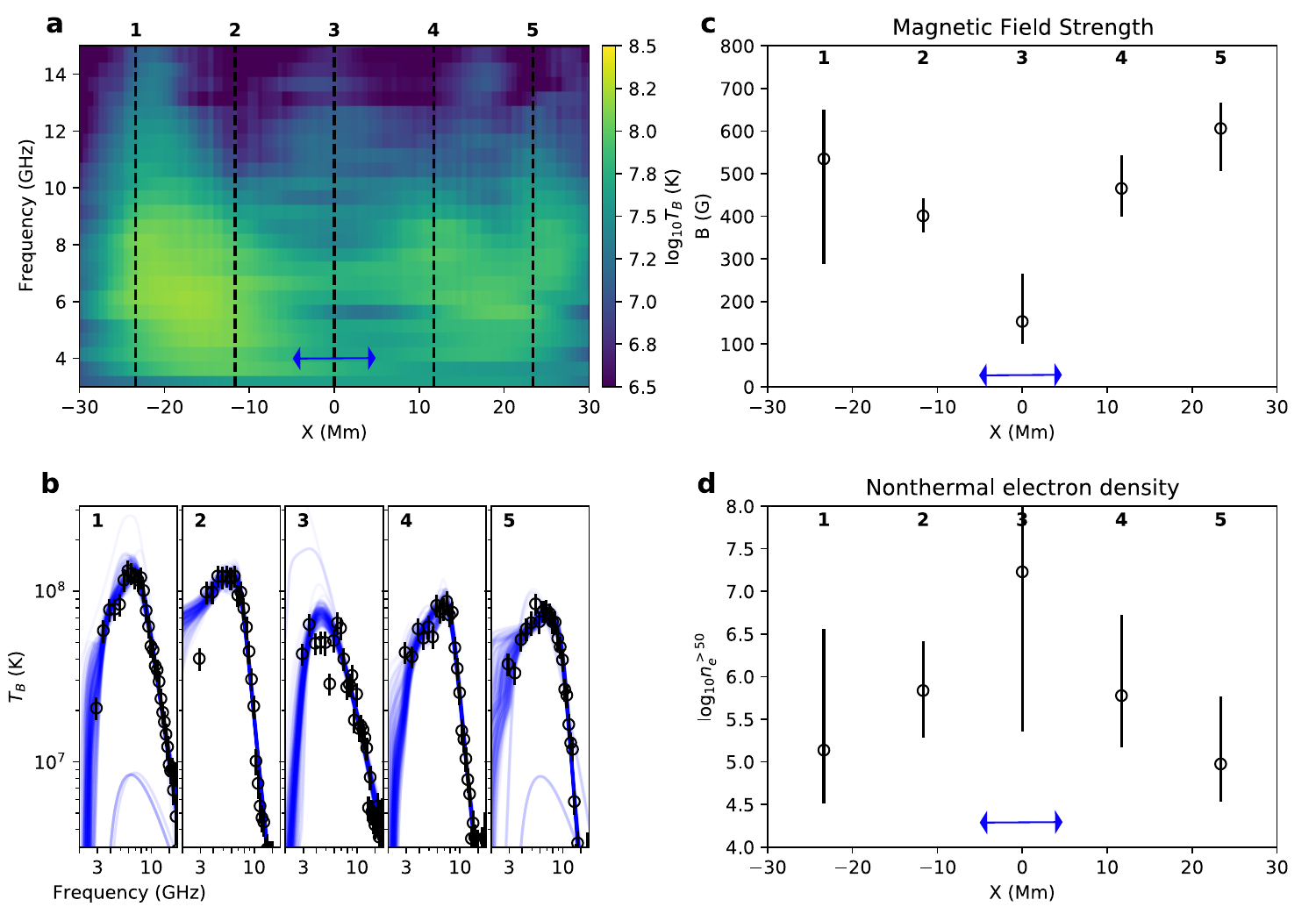}
\end{center}
\caption{Spatially resolved microwave spectra and derived source parameters along the above-the-loop-top, arcade-like microwave source. (a) Frequency--space spectrogram obtained along a fiducial slit drawn along the spine of the microwave source (white dashed curve in Figure~\ref{fig:overview}(d)). (b) Example microwave brightness temperature spectra at five sampled locations along the slit (vertical dashed lines in (a) marked from 1 to 5). Black circles with error bars are the measurements, and blue curves are a subset of 200 randomly selected model spectra from the MCMC runs, each of which has a total of 800,000 samples. (c) and (d) Distribution of the best-fit magnetic field strength $B$ and total nonthermal electron density above 50 keV $n_e^{>50}$ along the slit. The double-sided blue arrows mark the approximate location of the magnetic bottle region.}\label{fig:microwave_spec}
\end{figure*}

In the microwave images observed by the Expanded Owens Valley Solar Array (EOVSA; \citealt{Gary2018eovsa}) in 2.5--18 GHz, no counterpart of the bright loop-top X-ray source is present. Instead, it features an arcade-shaped source that bestrides above the bright EUV flare arcade (orange contours in Figure~\ref{fig:overview}). The arcade-shaped source can be clearly seen in microwave images at all frequencies, with subtle changes in their appearance (left panels in Figure~\ref{fig:microwave_images}). Interestingly, the low-frequency images ($\lesssim$6~GHz) display a small ``gap'' at the central location (i.e., near $x=0$ Mm) with a relatively lower brightness temperature. 
With EOVSA's multifrequency imaging capability, one can derive microwave spectra from different locations of the images and perform spectral analysis. In Figure~\ref{fig:microwave_spec}(a), we show a frequency--space spectrogram obtained along a fiducial slit drawn along the spine of the microwave arcade passing the ALT HXR source at $x\approx 0$ Mm (dashed curve in Figure~\ref{fig:overview}(d)). Such a spectrogram is akin to those obtained by slit spectrographs, with the microwave intensity along the slit ``dispersed'' in frequency over the vertical axis. Figure~\ref{fig:microwave_spec}(b) shows microwave brightness temperature spectra derived from five equally spaced locations along the slit. All spectra show a positive spectral slope at low frequencies and a negative slope at high frequencies with a peak brightness temperature of $\sim$100 MK, consistent with gyrosynchrotron emission produced by nonthermal electrons gyrating in the coronal magnetic field \citep[e.g.,][]{1982ApJ...259..350D}. The spectral properties are primarily sensitive to the parameters of the nonthermal electron distribution and the magnetic field in the source.

Following \citet{Chen2020NatAs}, we use the Markov Chain Monte Carlo (MCMC) method to analyze the observed spectra (see Appendix~B for details). Figure~\ref{fig:microwave_spec}(c) shows the distribution of the best-fit magnetic field strength $B$ along the slit. It displays a local minimum of $B$ at the center of the slit around $x\approx 0$ Mm, suggestive of a magnetic bottle structure colocated with the ALT HXR source. Meanwhile, despite the relatively large uncertainties, the distribution of the total nonthermal electron density above 50 keV $n_{e}^{>50}$ (Figure~\ref{fig:microwave_spec}(d)) indicates that the best-fit value in the central magnetic bottle region is 2 orders of magnitude greater than locations in the loop legs at $\lvert x \rvert > 20$ Mm, implying that microwave-emitting nonthermal electrons are strongly concentrated in the magnetic bottle region. The best-fit nonthermal electron density in the magnetic bottle region is also consistent with that returned from the X-ray spectral analysis within uncertainties.

\begin{figure*}[!ht]
\begin{center}
\includegraphics[width=0.8\textwidth]{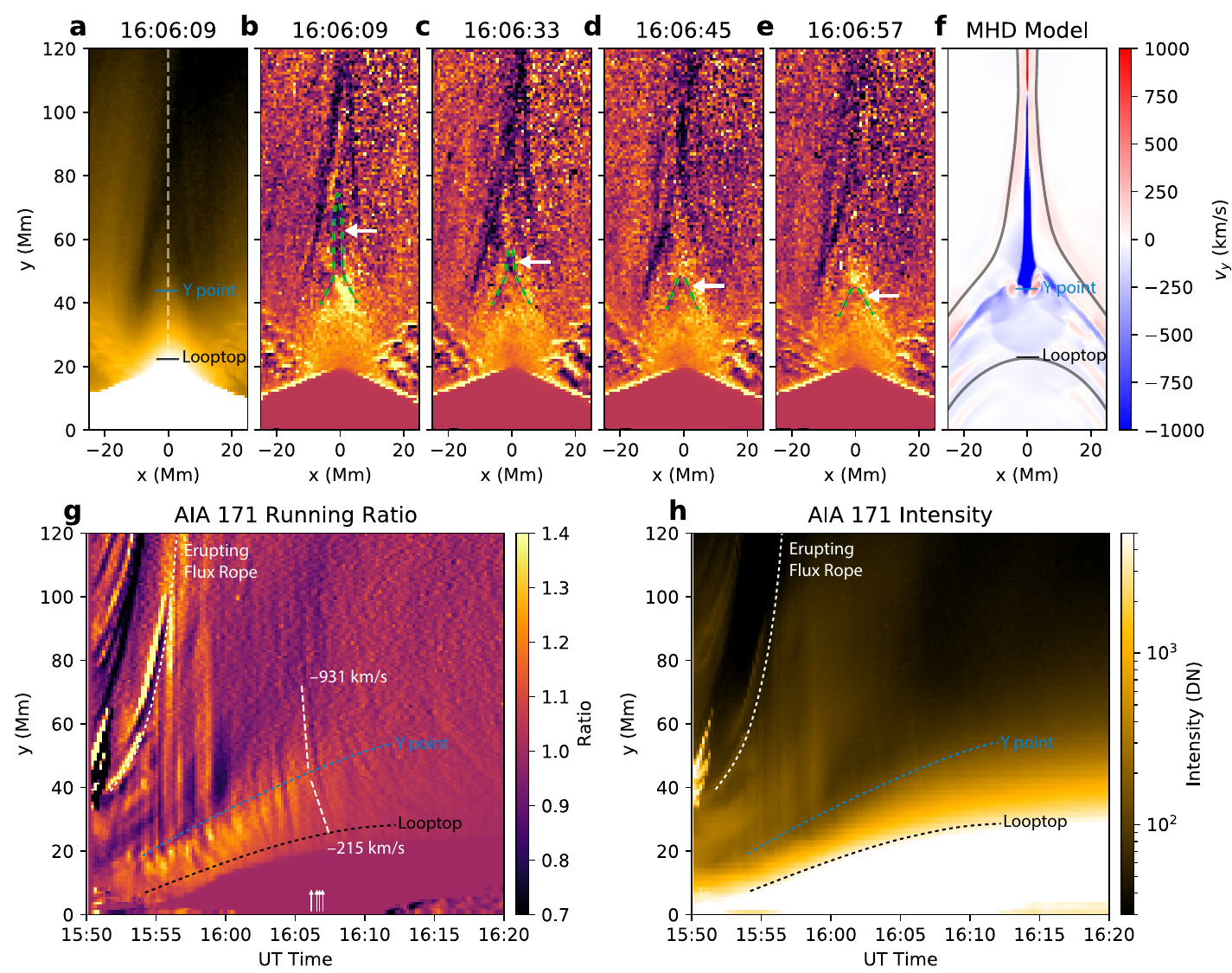}
\end{center}
\caption{Plasma outflows slow down and transform to a cusp shape as they pass the magnetic bottle hosting the Y point. (a) Context SDO/AIA 171 \AA\ EUV image at 16:06:09 UT. (b)--(e) Time series of running-ratio SDO/AIA 171 \AA\ images. Each plot shown is a ratio of the current image to a time 36 s earlier. The arrows indicate the downward plasma outflow that transforms from a linear shape to a cusp shape as it passes the Y point. (f) Corresponding MHD model showing the vertical component of the flow velocity. Downward plasma outflows (blue color) dominate the region above the Y point. (g) Time--distance plot of SDO/AIA 171 \AA\ running-ratio time-series images obtained along a vertical slice at the central current sheet region (white dashed line in (a)). Approximate tracks of the bottom of the erupting flux rope, Y point, and loop top are marked as dashed curves.  The example outflow event in (b)-(e) is marked by the two white dashed lines, which display an abrupt speed change after it passes the Y point. (h) Same as (g) but showing the original SDO/AIA 171 \AA\ intensity. }\label{fig:downflow}
\end{figure*}

\subsection{EUV Plasma Downflows}
In the ALT magnetic bottle region, the magnetic field lines associated with the downward reconnection outflows are expected to start from a nearly antiparallel configuration in the reconnection current sheet to a cusplike shape after they pass the Y point. Then, the outflows quickly slow down until they join the flare arcade and relax to a potential state with a looplike shape. This process is clearly shown in the SDO/AIA 171 Å EUV time-series images. In Figures~\ref{fig:downflow}(b)--(e), to better present the dynamic features, we plot a series of running-ratio images, each showing the intensity ratio of the current image to the one from 36 s earlier. 
There, a downward-moving feature at the central current sheet location is seen to transform from an initially linear shape to a cusp shape as it moves across a location at $y\approx 43$~Mm. Such a transformation is already suggestive of the presence of a Y point at this location. 

Furthermore, in Figure~\ref{fig:downflow}(g), we show a space-time diagram obtained at a vertical slice located at the central current sheet. Multiple downward-moving features can be distinguished as tracks that orient toward the lower right. The example downflow event shown in panels (b)--(e) has an initial speed of $>$900~km~s$^{-1}$ and quickly slows down to only $\approx200$~km~s$^{-1}$ after it passes the same location. Such a sudden slowdown motion of plasma downflows further supports the presence of the Y point. Coincidentally, the Y point identified with the EUV imaging observations alone matches almost exactly with the location of the ALT HXR source shown in Figure~\ref{fig:overview}. It is also fully consistent with the location of the Y-point-hosting magnetic bottle structure inferred from the microwave spectral imaging analysis, which has a weaker magnetic field strength. 

We note that the plane-of-sky-projected speed of the observed plasma downflow above the Y point ($\sim$900~km~s$^{-1}$) seems slower than the inferred local Alfv\'{e}n speed in the ALT magnetic bottle region ($v_{\rm A}\approx 1,900$~km~s$^{-1}$ for $B\approx 150$~G, constrained by our microwave spectral imaging analysis, and $n_{\rm th}\approx 3\times10^{10}$~cm$^{-3}$, constrained by Hinode/EIS measurements discussed in Appendix~C). By combining line-of-sight (LOS) flow measurements made by Hinode/EIS and plane-of-sky measurements from SDO/AIA, \citet{French2024} reported that despite that the reconnection current sheet appearing to be viewed perfectly edge-on, the nonzero LOS speeds in the downflow region suggest that the current sheet is tilted slightly away from the observer. Therefore, the physical flow speed should be greater than that measured in the plane of the sky, although the required projection correction is likely insignificant given the low LOS speed ($<$35~km~s$^{-1}$) measured by Hinode/EIS. Moreover, it has been shown in recent 3D modeling that the speed of the observed plasma outflows in EUV time-series images may be substantially slower than the intrinsic reconnection outflows \citep{Shen2023}. Lastly, as argued by \citet{Chen2020NatAs}, the low cadence of SDO/AIA (12 s) may simply render flows faster than $\sim$2000~km~s$^{-1}$ undetectable within the flaring region because they would traverse a large distance of $>$50 Mm in two frames. For these reasons, we argue that supermagnetosonic outflows may still be present to drive a fast-mode termination shock in the magnetic bottle region to facilitate the acceleration of the energetic particles. 

\section{MHD, Particle, and Emission Modeling}\label{sec:model}
\begin{figure*}[!ht]
\begin{center}
\includegraphics[width=0.9\textwidth]{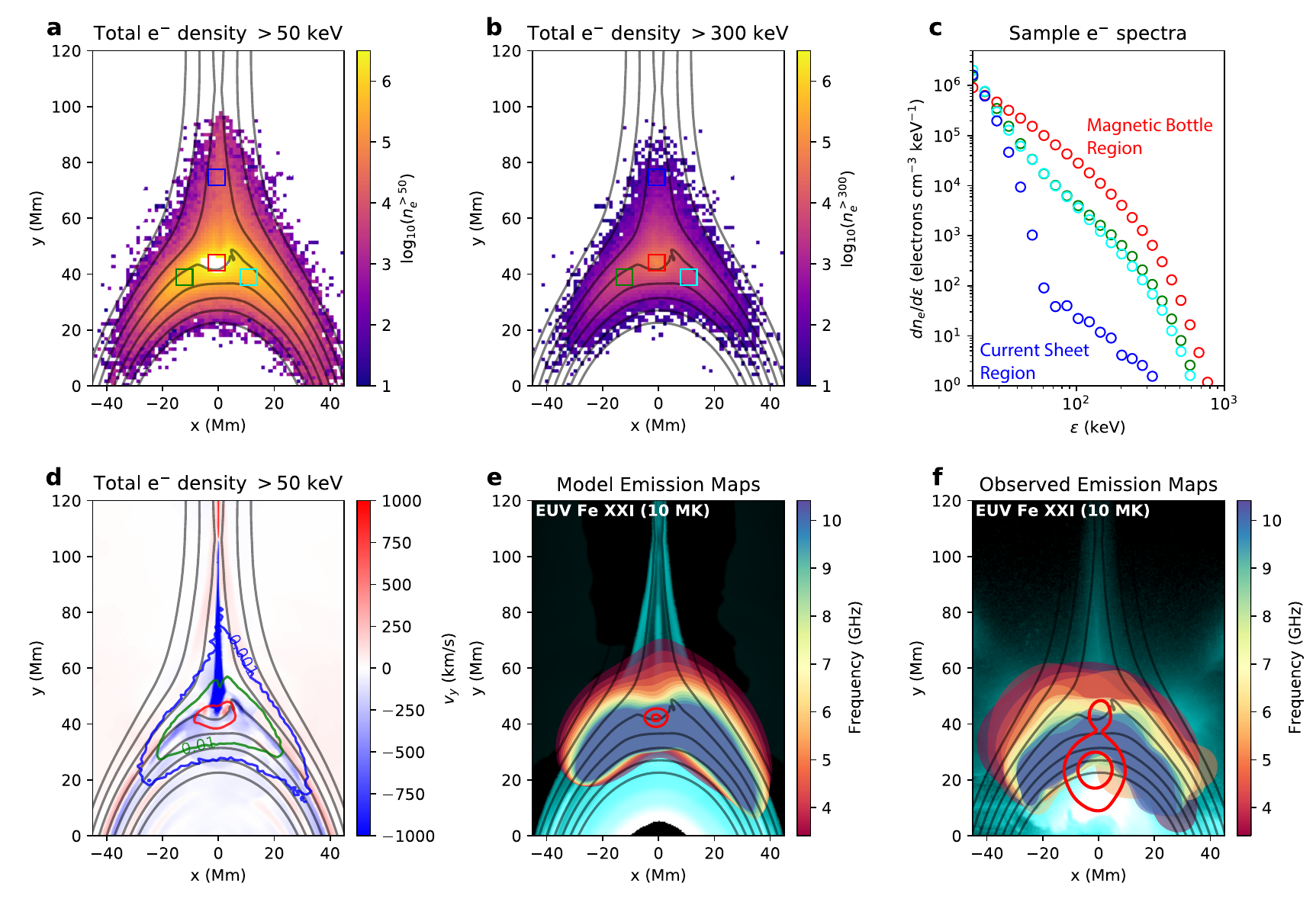}
\end{center}
\caption{Distribution of nonthermal electrons around the magnetic bottle region and the associated emissions. (a) and (b) Spatial distribution of $>$50 keV and $>$300 keV electrons, respectively. Note that they are shown on a logarithmic scale. (c) Example model nonthermal electron spectra from four regions marked in panels (a) and (b), with one located near the Y point (red), two in the loop legs (green and cyan), and one in the current sheet region (blue). (d) Contours of $>$50 keV electron distribution overlaid on the model $v_y$ map. Blue, green, and red contours are 0.1\%, 1\%, and 15\% of the peak electron density, respectively. In particular, the innermost red contour encloses 51\% of all the $>$50 keV electrons. (e) Synthetic SDO/AIA (background), 25--60 keV HXR (red contours), and multifrequency microwave sources (filled color contours) as calculated from the combined MHD and particle model. (f) Multifrequency observations identical to Figure~\ref{fig:overview}(d), overlaid with magnetic field lines derived from the model.}\label{fig:obs_model}
\end{figure*}

The multiwavelength analysis above suggests that the ALT magnetic bottle region is where the majority of the microwave- and X-ray-emitting nonthermal electrons are concentrated. To further elucidate the role of the magnetic bottle and the underlying physical processes, we combine a data-informed macroscopic MHD and particle model to simulate the acceleration and transport of energetic electrons in a realistic flare geometry. The resistive 2.5D MHD simulation we perform here is similar to those used in our earlier works \citep{Chen2015,Chen2019,Shen2018}, which features a dynamic fast-mode termination shock in the ALT magnetic bottle region. Appropriate scaling is applied to the MHD model to adjust to the observed flare size and observational constraints of the plasma properties. To model the electron acceleration and transport processes in the macroscopic flare geometry, we adopt the method used in \citet{Kong2019} by injecting pseudoelectrons into the MHD model and simulating their kinetic evolution by solving the Parker transport equation. In the model, the electrons are found to be primarily accelerated via the Fermi mechanism due to compression \citep{Fermi1949, Parker1965} in the magnetic bottle region, where the downward plasma flows collide head-on with the newly reconnected field lines. The fast-mode termination shock further facilitates the acceleration thanks to the sharp jump of physical parameters across the shock surface \citep{Kong2019}. We refer interested readers to Appendix~C and references therein for more technical details.

The combined MHD and particle model results in spatially, temporally, and spectrally resolved nonthermal electron and thermal plasma distribution in the simulation domain. Figures~\ref{fig:obs_model}(a) and (b) show the distribution of $>$50 keV and $>$300 keV nonthermal electrons, respectively, with example electron spectra derived from three different locations in the model shown in Figure~\ref{fig:obs_model}(c). After scaling the dimensionless electron distribution in the model with physical units constrained by the observations, similar to the microwave and HXR analysis results, the distribution of $>$50 keV electrons peaks in the magnetic bottle region with a number density of $5\times10^6$~cm$^{-3}$ and drops rapidly to $<$1\% of the peak value beyond $\lvert x \rvert > 20$ Mm (see contours in Figure~\ref{fig:obs_model}(d)).

\begin{figure*}[!ht]
\begin{center}
\includegraphics[width=0.9\textwidth]{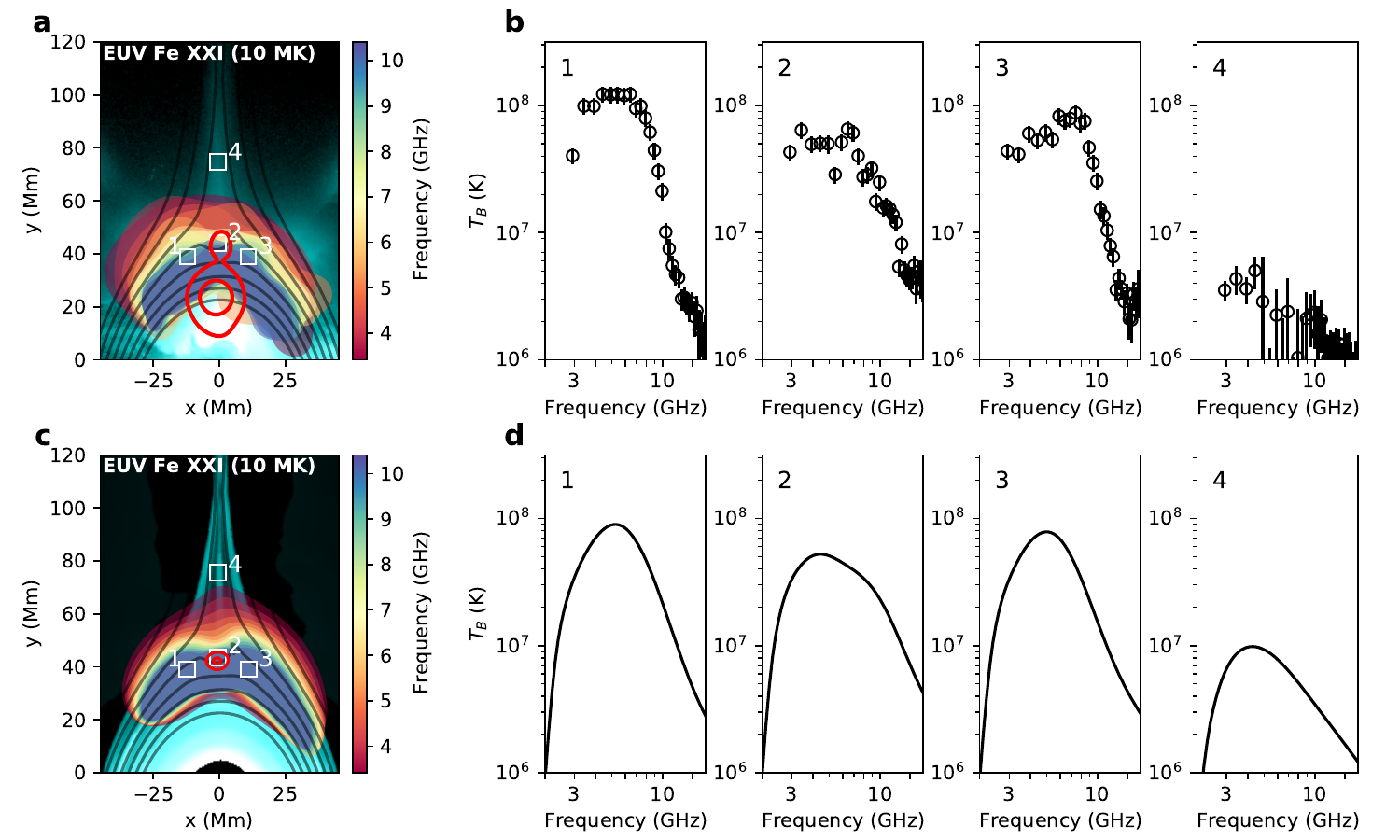}
\end{center}
\caption{Comparison between the observed and modeled microwave images and spectra. (a) Observed multifrequency microwave images overlaid on the SDO/AIA 131 \AA\ EUV image (identical to Figure~\ref{fig:obs_model}(f)). (b) Sample microwave spectra derived from four selected regions marked in (a). (c) and (d) are similar to (a) and (b) but instead show the modeled multifrequency microwave images and the corresponding sample spectra. }\label{fig:obs_model_spec}
\end{figure*}

By combining the resulting distribution of energetic electrons from the particle model $dn_{e}/d\varepsilon(x,y,\varepsilon)$ and plasma properties including magnetic field $B(x,y)$, thermal plasma density $n_{\rm th}(x,y)$, and temperature $T(x,y)$ from the MHD model, we generate synthetic SDO/AIA EUV, RHESSI HXR, and EOVSA microwave images at different energy/frequency bands. Appropriate emission mechanisms are assumed for each emission type: optically thin line emissions for EUV, thin-target bremsstrahlung for HXRs, and nonthermal gyrosynchrotron emission for microwaves. The corresponding instrument response is also considered (see Appendix~D for details). Figure~\ref{fig:obs_model}(e) shows a composite of the resulting synthetic SDO/AIA 131 \AA\ EUV image (background), RHESSI 25--60 keV HXR image (red contours), and multifrequency EOVSA microwave images (filled color contours). All of the images display a striking resemblance to the observations shown in Figure~\ref{fig:obs_model}(f): the EUV 131 \AA\ image displays a bright, closed arcade with a cusp-shaped top. A compact HXR source is located in the magnetic bottle region near the fast-mode termination shock above the flare arcade. Meanwhile, similar to the observations, the multifrequency microwave source resembles an arcade-like shape at the outer rim of the flare arcade. Figure~\ref{fig:microwave_images} shows a side-by-side comparison between the observed and modeled microwave images at the same frequencies. Figures~\ref{fig:obs_model}(e) and (f) display another comparison with the microwave images shown as overlaid filled contours colored from red to blue for increasing frequencies. One can see that the model images resemble the observed ones remarkably well. Moreover, as shown in Figure~\ref{fig:obs_model_spec}, not only does the modeled microwave source morphology achieve an excellent match with the observations, but the spatially resolved microwave spectra derived from different regions also yield a qualitative agreement, demonstrating the success of our approach. The microwave source's appearance is very different from its HXR counterpart mainly because, unlike the X-ray bremsstrahlung, gyrosynchrotron radiation has a strong dependence on the local magnetic field.

Intriguingly, unlike the HXR source that is highly localized at the ALT magnetic bottle, similar to the observations, the multifrequency microwave source extends well beyond the region toward the directions of the current sheet and the two footpoints. 
Such a large extension suggests that, while the majority of the energetic electrons are confined within the ALT HXR source near the Y point---in the model, it contains $\sim$50\% of the total $>$50 keV electrons (see red contour in Figure~\ref{fig:obs_model}(d))---a small fraction of these electrons manage to escape and spread beyond the magnetic bottle region (albeit ``diluted'' to a much smaller density), giving rise to the extended microwave source with an arcade-like shape.

We note that in the observations, another strong loop-top X-ray source is present, whose centroid is located at $y\approx 24$ Mm (Figure~\ref{fig:obs_model}(f)). Spectral analysis results suggest that it is extremely hot (15--27 MK) and dense (up to $10^{12}$~cm$^{-3}$). This source has been interpreted as the result of thermalization of previously accelerated nonthermal particles \citep{Veronig2004}, colliding chromospheric upflows \citep{Reeves2007}, or compression by slow shocks formed by sheared reconnection \citep{Longcope2011looptop}. Reproducing this additional source in a model requires the inclusion of feedback between the thermal plasma and nonthermal particles, more accurate treatment of the chromospheric evaporation processes, and modeling of the reconnection processes in the third dimension, which is beyond the scope of our current work. 

\section{Discussion and Conclusion}\label{sec:discussion}
We have demonstrated that our model, which includes both MHD and particle processes in a realistic flare geometry, can reproduce emission signatures that are well matched to multiwavelength observations. In the model, the energetic electrons are found to be primarily accelerated in the magnetic bottle region by the converging flows and facilitated by the fast-mode termination shock via the Fermi mechanism. The accelerated electrons are further trapped there due to pitch-angle scattering by turbulence. This is the first time that synthetic observables in both the thermal and nonthermal regimes have been generated from a self-consistent, macroscopic numerical model to compare with microwave, EUV, and X-ray imaging and spectroscopy observations. Such a remarkable agreement between the modeled and observed emissions suggests that this model of electron acceleration and transport is a viable approach during this period of interest. 

The presence of the magnetic bottle structure in the ALT region of this event is well supported by its lower magnetic field strength derived from the microwave imaging spectroscopy data (Figure~\ref{fig:microwave_spec}), as well as the observed abrupt change in the morphology and speed of the EUV plasma downflows (Figure~\ref{fig:downflow}). Direct evidence for fast-mode termination shocks is more elusive due to various challenges in identifying them observationally (see, e.g., discussions in \citealt{Chen2019}). However, recently, new and convincing evidence for their existence started to emerge. For this particular event, by using EUV imaging spectroscopy data recorded by Hinode/EIS, \citet{French2024} reported a sharp gradient in the Doppler velocity of hot ($\sim$18 MK) EUV downflows in the ALT region. The location and characteristics of the sharp velocity gradient agree very well with an MHD model that features a fast-mode termination shock at the same location. In addition, \citet{Cai2019} argued that the hot ``supra-arcade fan'' structure observed by SDO/AIA, Hinode/EIS, and the Interface Region Imaging Spectrograph was possibly caused by a termination shock, albeit the reported structure is not located above the main flaring arcade but in a loop system south of that.

\begin{figure*}[!ht]
\begin{center}
\includegraphics[width=0.9\textwidth]{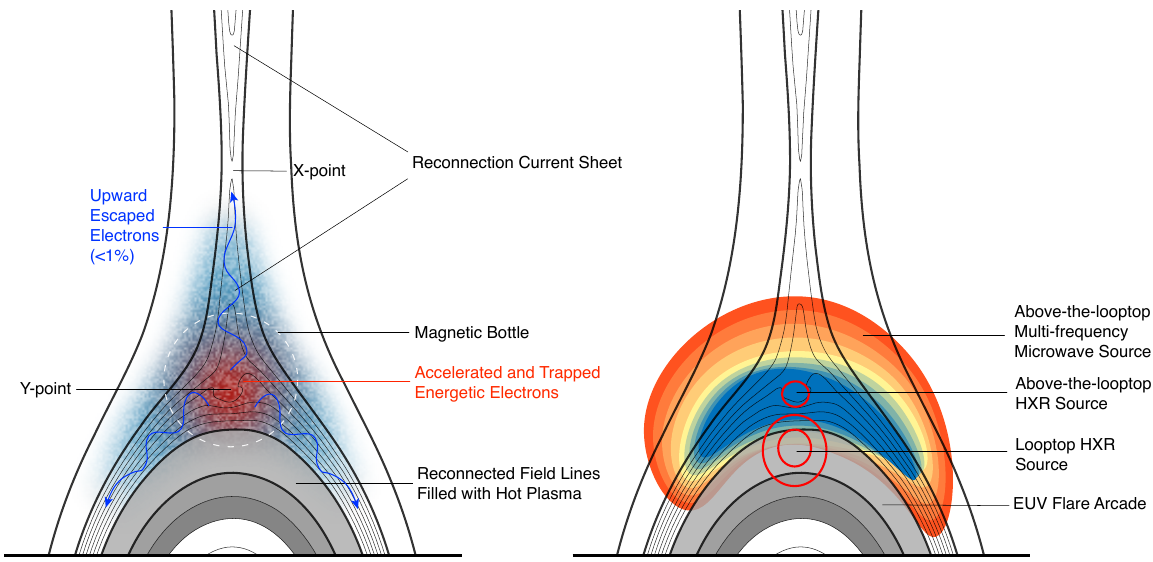}
\end{center}
\caption{Physical picture suggested by our combined observation--modeling results. Energetic electrons are accelerated via the Fermi mechanism and trapped in the ALT magnetic bottle region due to turbulence. The reconnected and relaxed magnetic loops filled with hot plasma are observed as a bright EUV flare arcade with a hot loop-top HXR source. The highly concentrated energetic electrons near the Y point produce the ALT HXR source via bremsstrahlung. Meanwhile, a small fraction of energetic electrons escaping from the magnetic bottle region produces the observed multifrequency microwave source with a large extension, which has an arcade-like shape bestriding above the EUV flare arcade. }\label{fig:cartoon}
\end{figure*}

The escaped nonthermal electrons from the ALT region gyrate in the magnetic field and produce microwave sources that form an arcade-like shape bestriding \textit{above} the bright EUV flare arcade. The absence of an HXR footpoint source in our case suggests that it is either partially occulted by the solar limb at the time or the flux of nonthermal electrons reaching the footpoints is insufficient to produce observable HXR footpoint emission in the presence of a bright loop-top source. In other cases where HXR footpoint sources are co-observed, they are often found to be located at the outer edge of bright EUV flare arcades \citep[e.g.,][]{LiuR2013,Krucker2014}. These observations nicely corroborate our scenario: they result from accelerated nonthermal electrons escaping from the ALT magnetic bottle region and reaching the footpoints along the freshly reconnected field lines \citep[see, e.g., recent modeling results by][]{Kong2022}.

As depicted in Figure~\ref{fig:cartoon}, most of the accelerated electrons are confined near the Y point, producing the observed ALT HXR source. Meanwhile, the full extension of the observed microwave source traces 0.1\%--1\% of the peak $>$50 keV nonthermal electron density, or merely a millionth of the background plasma density! Our observations demonstrate the extreme sensitivity of microwave emission to even a tiny population of flare-accelerated nonthermal electrons. Such a sensitivity partially benefits from the Razin effect \citep{Ginzburg1965}. This effect strongly suppresses the microwave brightness in regions with relatively low magnetic field strength and high thermal plasma density, which are the exact features of the magnetic bottle region in our case. Strong Razin suppression occurs below a critical frequency of $\nu_{c}\approx 20n_{\rm th}/B$, or $\approx$5--6~GHz in the magnetic bottle region. This suppression effect is likely responsible for the apparent ``gap'' in the observed and modeled microwave sources at around $x=0$ Mm for frequencies $\lesssim$6 GHz (see Figure~\ref{fig:microwave_images}). Also, it effectively reduces the microwave brightness of the magnetic bottle region despite its high concentration of nonthermal electrons and, in turn, facilitates the detection of microwave sources arising from an extremely small nonthermal electron population outside the region.

The strong concentration of nonthermal electrons within the magnetic bottle region has important implications for electron acceleration and transport processes. First, it is highly likely that the nonthermal electrons are primarily accelerated in the magnetic bottle region itself rather than injected from elsewhere, such as high up in the reconnection current sheet or low in the loop legs. As shown in Figure~\ref{fig:obs_model}(d), our joint observation--modeling results suggest that the nonthermal electron density rapidly drops to $<$1\% of the peak value outside the ALT magnetic bottle region in all directions. Such a large density contrast implies that if the injection-and-trapping process is responsible, it must be a slow process. For instance, assuming all the electrons are injected from the upper current sheet, pure injection of 50 keV electrons without any loss takes $\tau^{\rm inj}\approx N_e^{\rm ALT}/F_e^{\rm inj} = n_e^{\rm ALT} L_y^{\rm ALT}A_{xz}^{\rm ALT}/(n_e^{\rm CS}v^{\rm inj}A_{xz}^{\rm CS})\approx (n_e^{\rm ALT}/n_e^{\rm CS})(A_{xz}^{\rm ALT}/A_{xz}^{\rm CS})L_y^{\rm ALT}/v^{\rm inj}$, where $n_e^{\rm ALT}/n_e^{\rm CS}>100$ is the ratio of the nonthermal electron density between the ALT and current sheet region, $A_{xz}^{\rm ALT}/A_{xz}^{\rm CS}>50$ is the expansion factor of the cross section between the magnetic bottle region (with a width of $L_x^{\rm ALT}\approx10$ Mm) and the narrow current sheet (with a width of $L_x^{\rm CS}\approx0.2$ Mm, the grid size of the numerical model), $L_y^{\rm ALT}$ is the vertical extension of the ALT region taken to be $\sim$5 Mm as suggested in the model, and $v^{\rm inj}$ is the injection speed of the electrons, taking the kinetic speed of 50 keV electrons $v_e^{50}\approx0.41c$ as the upper limit. The estimated injection timescale to produce the ALT source is $\tau_{\rm inj}>200$ s, even without any loss. This timescale is much longer than the typical acceleration timescale inferred from HXR and radio emissions, which can display rapid fluctuations at second or even subsecond scales (e.g., \citealt{Fletcher2011} and references therein), although we note that the acceleration in the flare gradual phase may be less variable and, as such, may have an inherently longer timescale. While we cannot entirely rule out such an injection-and-trapping scenario owing to the insufficient observational constraints for the precise dimensions of the ALT region (particularly in the LOS $z$-direction), we conclude that local acceleration and trapping in this region is a more likely scenario to account for the profound concentration of the nonthermal electrons in the ALT magnetic bottle region.

Second, such a concentration requires effective trapping of the energetic electrons. In our model, we invoke diffusion in the strong pitch-angle scattering regime induced by turbulence with a prescribed Kolmogorov-type spectrum. In this case, pitch-angle diffusion quickly leads to an isotropization of the electron distribution, and the transport processes can be approximated by Parker's transport equation. Although our model yields a good match with the observations after adjusting for the (essentially unknown) diffusion parameters, it is certainly not a unique approach. For example, other analytical and numerical models for the trapping-and-precipitation processes can be found in the literature, some of which involve treatments for the diffusion processes due to both momentum and pitch-angle scattering as well as collisional loss \citep{ChenQ2013,Kontar2014,Kong2022}. 

Finally, our results provide new insights into understanding the puzzling departure from equipartition between the upward-escaped and downward-retained energetic electron population reported by studies that combine remote-sensing and \textit{in situ} observations, which have concluded consistently that only 0.1--1\% of the flare-accelerated energetic electrons manage to escape to interplanetary space \citep{Lin1974,Krucker2007,Wang2021,WangM2023}. Such a large imbalance not only poses a challenge for understanding the particle acceleration and transport processes but also has important implications for space weather in solar and extrasolar systems. 
In our scenario, this imbalance naturally occurs because the primary acceleration site is the magnetic bottle located \textit{below} the X point, with the newly reconnected, cusp-shaped field lines in the large-scale current sheet acting as a nearly closed structure to limit the upward-directed electrons from escaping (Figure~\ref{fig:cartoon}). Along with efficient trapping, as in our case, the energetic electron density reaching the X point can be only $<$1\% of that at the core acceleration region, giving rise to an extremely small fraction of these energetic electrons that enter interplanetary space. 

We note, however, that since our model does not involve an exhaustive search in the parameter space and does not include all possible particle energization/transport mechanisms, it is by no means exclusive. Other scenarios that result in an efficient energization and confinement of nonthermal electrons in the magnetic bottle region while allowing only a small fraction of them to escape may also be possible. Some candidates may include magnetic islands \citep{Drake2006, Guidoni2022}, collapsing traps \citep{1997ApJ...485..859S}, and intense shock heating \citep{1994Natur.371..495M, Mann2024}, yet rigorous data--model comparisons are required to further examine these models. Moreover, in order to achieve a more definitive understanding, extensive studies are needed for a large sample of flare events with different intensities and geometries. Last but not least, to make further progress, next-generation telescopes capable of performing radio and HXR imaging spectroscopy with orders-of-magnitude higher dynamic range and sensitivity, as well as more sophisticated models, are desired. Notable future telescope concepts include the Frequency Agile Solar Radiotelescope \citep{Chen2023fasr,Gary2023fasr} in radio wavelengths and the Focusing Optics X-ray Solar Imager \citep{Christe2016foxsi} or its variants (e.g., COMPLETE; \citealt{Caspi2023}) in X-rays.


\acknowledgements

This work is primarily funded by NASA Heliophysics Supporting Research grant 80NSSC20K1318 to NJIT, SAO, UMN, and LANL. B.C. and S.Y. received additional support from NSF grants AST-1735405 and AST-2108853 to NJIT. The authors are grateful to Dr. Harry Warren for providing the Hinode/EIS temperature and emission measure maps. They also thank Drs. Alexey Kuznetsov and Gregory Fleishman for making their fast gyrosynchrotron codes available. X.K. is supported by NSFC grants 42074203 and 11873036. C.S. is supported by NASA grants 80NSSC19K0853 and 80NSSC21K2044 and NSF AST 2108438 to SAO. X.L. acknowledges the support from NASA through grant 80NSSC21K1313, NSF grant AST-2107745, SAO through subcontract SV1-21012, and LANL through subcontract No. 622828. The Expanded Owens Valley Solar Array (EOVSA) was designed and built and is now operated by the New Jersey Institute of Technology (NJIT) as a community facility. EOVSA operations are supported by NSF grant AGS-2130832 and NASA grant 80NSSC20K0026 to NJIT. The authors are grateful to the RHESSI, SDO, SOHO/LASCO, and GOES-R/SUVI teams for making their data publicly available.

\facilities{OVRO:SA, RHESSI, SDO, Hinode, SOHO, GOES}

\section*{Appendix A: X-ray data analysis}\label{append sec:X-ray}
\begin{figure*}[!ht]
\begin{center}
\includegraphics[width=1.0\textwidth]{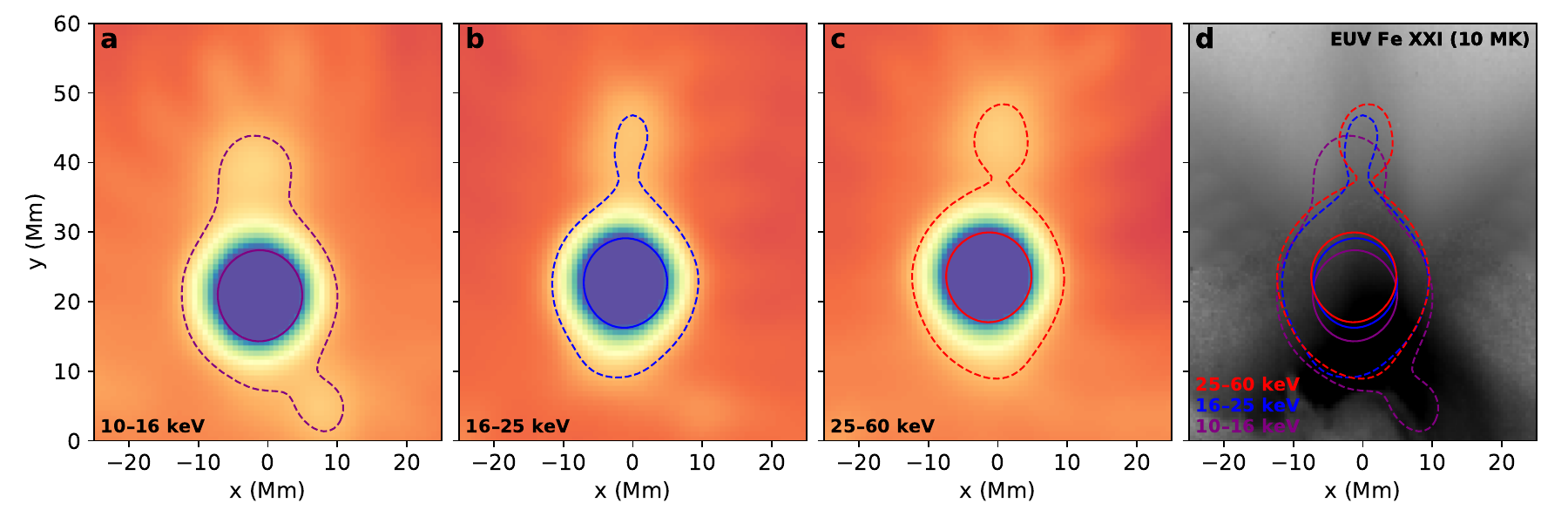}
\end{center}
\caption{RHESSI X-ray images for the time of interest on 2017 September 10. The time interval used for imaging is 16:10:00--16:11:10 UT. The energy ranges used for imaging are 10--16 keV, 16--25 keV, and 25--60 keV, shown in panels (a), (b), and (c), respectively. The solid and dashed lines denote the 50\% and 10\% contours. Panel (d) shows X-ray images of all three energy ranges overlaid on the SDO/AIA 131 \AA\ (\ion{Fe}{21}) EUV image at the closest time (shown in reversed gray scale). }\label{fig:xray_img}
\end{figure*}

RHESSI observed the event with its detectors 1, 3, 6, and 8. During the time of interest, the attenuator state was set to A3, meaning both the thin and thick attentuators were used. X-ray imaging is performed using measurements made by detectors 3, 6, and 8 with the CLEAN algorithm \citep{Hurford2002}. The nominal angular resolution of the finest grid used for imaging (detector 3) is $6''.8$, which is used as the FWHM width of the synthesized beam for restoring the CLEAN images. Figure~\ref{fig:xray_img} shows the resulting X-ray images integrated between 16:10:00 UT and 16:11:10 UT in 10--16 keV, 16--25 keV, and 25--60 keV.

X-ray spectral analysis is performed using the \texttt{ospex} software \citep{Schwartz2002}. We use the data from detector 3 (which had the lowest sensitivity at the time and hence was less affected by the pileup effect) to obtain the X-ray count flux spectrum between 16:10:00 UT and 16:11:08 UT. The resulting X-ray photon flux spectrum after applying the instrument response matrix is shown in Figure~\ref{fig:xray_spec} as a black curve. Spectral modeling is performed in the 12--60 keV range using two isothermal functions and one nonthermal bremsstrahlung model. Also included in the spectral modeling is the pileup module. We performed spectral analysis assuming both the thick- and thin-target regimes. In both regimes, the thermal components, which dominate the spectrum below $\sim$30 keV, include a hot $\sim$15 MK loop-top source with a volume emission measure of $\approx 2\times10^{51}\,\mathrm{cm}^{-3}$. A secondary superhot component with a temperature of 27--28 MK and a volume emission measure of $\approx 4\times10^{49}\,\mathrm{cm}^{-3}$ is also present. Assuming a source volume $V\approx (10\,\mathrm{Mm})^3$ according to the size of the source shown in the X-ray image, the plasma density for the two thermal components becomes $\approx1\times10^{12}$~cm$^{-3}$ and $2\times10^{11}$~cm$^{-3}$, respectively. Such a high density associated with the hot 15 MK component is sufficient to stop all nonthermal electrons up to 70 keV in the source (with a half-width of $\sim$5 Mm) through Coulomb collisions \citep{Tandberg-Hanssen1988}.

For the nonthermal component, if the component falls into the thick-target bremsstrahlung regime due to interactions with an extremely dense plasma environment, such as the loop-top X-ray source with a density of $n_{\rm th}>10^{12}$~cm$^{-3}$, the best fit yields a nonthermal electron distribution with a total electron flux of $F_{e}^{>50}\approx 5\times10^{33}$~electrons~s$^{-1}$ above 50 keV and a power-law index of the electron flux spectrum of $\delta^{\rm thick}\approx 5.3$. However, if the component is instead associated with thin-target bremsstrahlung in a relatively tenuous coronal plasma environment, such as the ALT region with a density of $n_{\rm th}\approx3\times10^{10}$~cm$^{-3}$, the fit returns a normalization factor of $n_{\rm th}Vf_e^{>50} = n_{\rm th}L_zF_e^{>50}=1.6\times 10^{54}\,\mathrm{cm}^{-2}\,\mathrm{s}^{-1}$ with a power-law index of $\delta^{\rm thin}=4.7$, where $L_z$ is the source column depth, $f_e^{>50}=\int v_e(\varepsilon)\frac{dn_e}{d\varepsilon}  d\varepsilon$ is the total electron flux density above 50 keV, and $F_e^{>50} = f_e^{>50} A$ is the total $>$50 keV electron flux in a source cross section of $A$. Taking a background plasma density of $n_{\rm th}\approx3\times10^{10}$~cm$^{-3}$ and a source column depth of $L_z\approx 10\,\mathrm{Mm}$, the nonthermal electron flux above 50 keV is one order of magnitude greater than the thick-target case, at $F_{e}^{>50}\approx 5\times10^{34}$~electrons~s$^{-1}$. The corresponding total nonthermal electron density above 50 keV is $n_e^{>50}\approx F_{e}^{>50}/(v_{e}^{50}A)\approx 4\times10^{6}$~cm$^{-3}$. We find that the parameters returned from the thin-target regime may yield a better agreement with the microwave analysis results.

As RHESSI was approaching the end of its operations at the time of the observation, only four of the nine detectors operated nominally, which limited its imaging capabilities. Furthermore, SOL2017-09-10 was one of the brightest flares ever observed by RHESSI. Despite both attenuators being inserted, the count rate nevertheless stayed high, and pileup effects occurred (i.e., two photons arrive essentially simultaneously, and they are therefore measured as a single photon with the summed energy of the two individual photons). While the time used for the analysis is $\sim$12 minutes past the flare peak, which has lessened the pileup issue, and pileup correction following the standard procedure has been applied in our fitting, the spectral analysis is possibly still pushing to the limits of the instrumentation.  Hence, despite the fact that the spectral fitting results are consistent with other complementary data, we suggest that they should be considered with some level of caution. For imaging, pileup correction is not available, making the exact partition of the X-ray photons between the loop top and ALT X-ray source undetermined.  
However, as pileup only affects imaging by removing a rather small fraction of photons as they pile up and appear at higher energies, the imaging morphology is not much affected for energies below $\sim$36 keV (2 times the peak of the count spectrum at $\sim$18 keV). 
We therefore conclude that the double-source structure seen in the X-ray images is trustworthy. However, we cannot draw any firm conclusion on the true shape and brightness of the ALT X-ray source. 

\section*{Appendix B: Microwave data analysis}\label{append sec:microwave}
EOVSA observed the SOL2017-09-10 X8.2 flare event in the range 2.5–18 GHz with 31 evenly spaced spectral windows, each of which has a bandwidth of 160 MHz. The center frequencies of these spectral windows are $\nu = 2.92 + 0.5n$ GHz, where $n$ is the spectral window number from 0 to 30. Different from \citet{Chen2020NatAs} and \citet{Fleishman2022}, which analyzed the earlier and main impulsive phase of the event, respectively, the time of interest for this study is around 16:10 UT, during the gradual phase of the flare when the magnetic flux rope has already propagated to a large distance, allowing detailed observation--modeling comparison. Methods used for calibrating and imaging the EOVSA data are identical to our earlier studies (see \citealt{Gary2018eovsa} and \citealt{Chen2020NatAs}). The integration time used for synthesis imaging is 4 s, and a circular beam with an FWHM size of $10''.2/[\nu/{10\ \rm GHz}]$ is used to restore the final microwave images after deconvolution. Figure~\ref{fig:microwave_images} shows an example of multifrequency EOVSA images from 3.4 GHz ($n=1$) to 10.4 GHz ($n=15$) at 16:10:36 UT.

Each spatially resolved microwave spectrum used for spectral analysis is derived from the average brightness temperature in an $8''\times8''$ area (corresponding to the resolution at $\sim$12.4 GHz). For each spatially resolved spectrum, a power-law nonthermal electron distribution $dn_e(\varepsilon)/d\varepsilon$ with a spectral index of $\delta'$ and total nonthermal density $n_{\rm nth}^{>50}$ above 50 keV is used to model the spectrum based on the fast gyrosynchrotron codes \citep{Fleishman2010}. Other model parameters used in the fit include the magnetic field strength $B$, thermal plasma density $n_{\rm th}$ of the source, and the viewing angle $\theta$ with respect to the magnetic field direction. The low- and high-energy cutoff of the electron distribution $\varepsilon_{\rm min}$, $\varepsilon_{\rm max}$, and the plasma temperature are fixed to 25 keV, 10 MeV, and 15 MK, respectively. For the MCMC analysis, similar to \citet{Chen2020NatAs}, we use a logarithmic likelihood function in the following form:
\begin{equation}
    \ln p = -\frac{1}{2}\sum_n \left[(T_{b, i}^{\rm o} - T_{b, i}^{\rm m})^2/\sigma_{T_{b,i}}^2 + \ln (2\pi\sigma_{T_{b,i}}^2)\right],
\end{equation}
where $T_{b,i}^{\rm o}$ and $T_{b,i}^{\rm m}$ are the observed and modeled brightness temperature at frequency $\nu_i$, respectively, and $\sigma_{T_{b,i}}$ is the corresponding uncertainty estimated by adding the rms brightness temperature of a region in the image without any sources and a fractional error in the source brightness temperature (assumed to be 15\%) in quadrature. We then use \texttt{emcee} \citep{2013PASP..125..306F}, a Python implementation of the affine-invariant MCMC ensemble sampler \citep{2010CAMCS...5...65G}, to sample the parameter space according to the likelihood function. The multiparameter posterior distribution allows us to find the best-fit model parameter values, taken as the median value of the samples in each marginalized distribution (i.e., the 50th percentile). The lower and upper 1-$\sigma$ uncertainties are taken as the 16th and 84th percentiles of the samples in each distribution, respectively. For the MCMC analysis of each spectrum, we use 100 ``walkers'' to sample the parameter space, each of which draws a total of 8000 samples. 

\section*{Appendix C: MHD and Particle Modeling} \label{append sec:model}
The setup of the MHD model follows similar procedures as those described in \citet{Shen2018} but with a different scaling to better match the observations of the particular event of interest. The initial setup is a Harris-type vertical current sheet centered at $x=0$ that separates two regions of the magnetic field with opposite polarity, which are line-tied at the lower boundary. A guide field of $B_{g} = 0.1B_0$ is introduced (where $B_0$ is the normalized magnetic field), and the initial background plasma beta is set to $\beta_0 = 0.01$. The magnetic field lines are line-tied to the bottom boundary and are open at the top boundary. Reconnection proceeds in the current sheet and forms a series of postflare arcades. Above the loop top, reconnection continues in the current sheet, driving the flare evolution and plasma dynamics. For thermodynamic treatment, classical Spitzer thermal conduction is used. In this study, we focus on a period in the MHD simulation (96.5--97.5$t_0$ in \citealt{Shen2018}) when the reconnection outflow is mostly laminar (i.e., without plasmoids), and the associated fast-mode termination shock is well defined and nearly symmetric. 

\begin{figure*}[!ht]
\begin{center}
\includegraphics[width=1.0\textwidth]{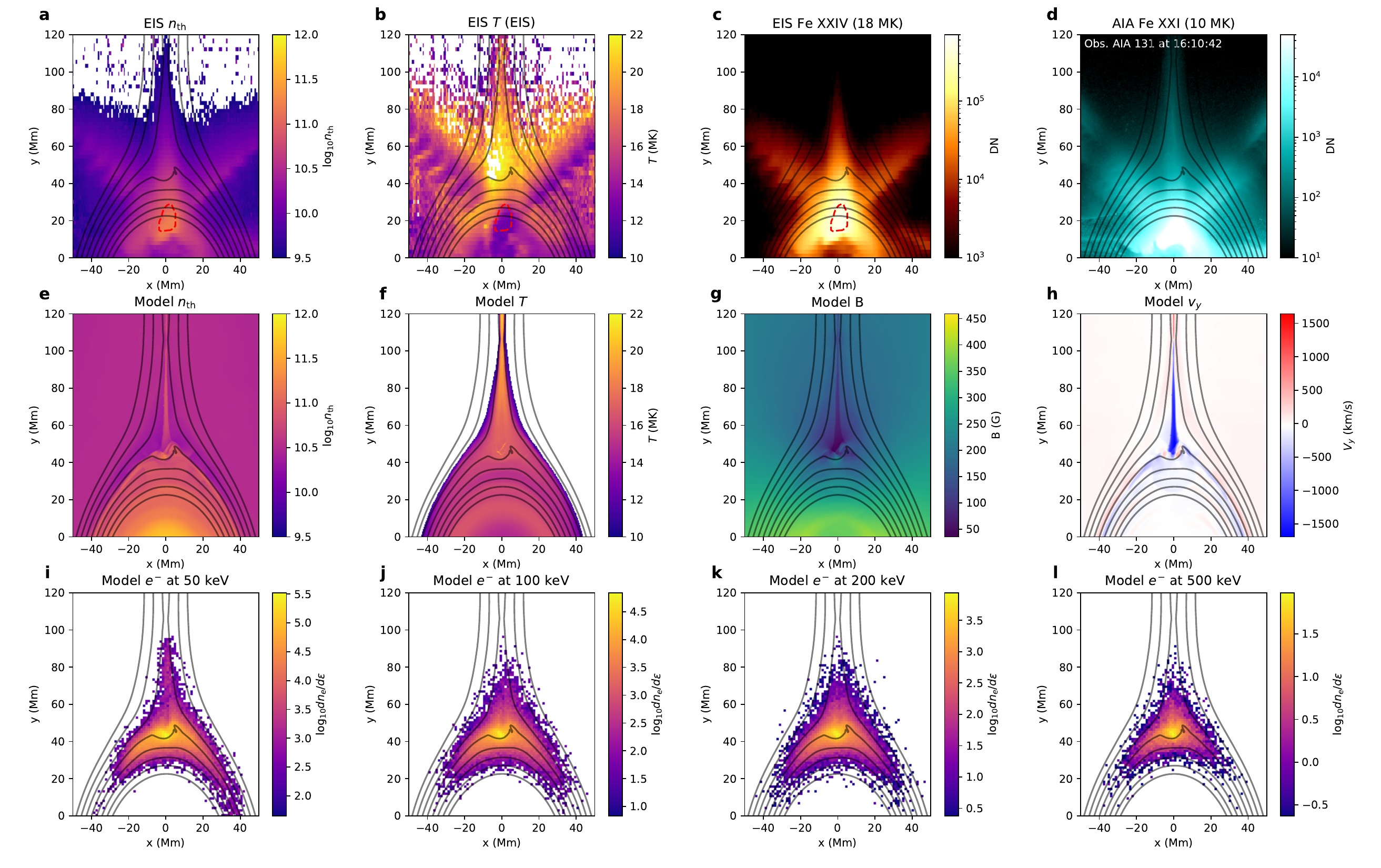}
\end{center}
\caption{Data-informed MHD and particle modeling of the 2017 September 10 flare during its gradual phase at 16:10 UT. (a) and (b) Plasma density and temperature maps derived from Hinode/EIS measurements of the \ion{Fe}{24} line pairs. (c) and (d) Hinode/EIS Fe \ion{Fe}{24} and SDO/AIA 131 \AA\ (\ion{Fe}{21}) intensity maps, which have a peak sensitivity to 18 MK and 10 MK plasma, respectively. The red contours outline the regions where the EIS intensity is saturated and has unreliable temperature/density diagnostics. Also, note the ``X''-shaped artificial diffraction pattern in the EIS and AIA images. (e)--(h) Distribution of plasma density $n_{\rm th}$, temperature $T$, magnetic field strength $B$, and vertical component of the plasma velocity $v_y$ from the MHD model. (i)--(l) Distribution of nonthermal electrons at 50 keV, 100 keV, 200 keV, and 500 keV, respectively, from the particle model. Black curves in all panels are magnetic field lines derived from the MHD model. }\label{fig:model}
\end{figure*}

Since the MHD simulation is performed in dimensionless units, we can scale the model parameters to those constrained by our observations. In particular, the spatial scaling is done by comparing the sizes of the flare arcade and plasma sheet to the observations. The magnetic field scaling is informed by the microwave spectral diagnostics described in Appendix~B. Scaling of the thermal plasma density and temperature is mainly based on Hinode/EIS measurements of the flare arcade and plasma sheet regions with EUV imaging spectroscopy (see \citealt{Warren2018cs} for details). 
Figure~\ref{fig:model} shows distributions of plasma parameters in the MHD model at 97.4$t_0$, which include plasma density (panel (e)), temperature (panel (f)), magnetic field strength (panel (g)), and the vertical component of the plasma velocity (panel (h)). In comparison, we also show the plasma density and temperature maps derived from Hinode/EIS measurements of the \ion{Fe}{24} line pairs in panels (a) and (b)\footnote{Hinode/EIS analysis returns the emission measure. A uniform column depth of 10 Mm is assumed to estimate the plasma density in the flaring region.}, as well as the Hinode/EIS \ion{Fe}{24} and SDO/AIA 131 \AA\ (\ion{Fe}{21}) intensity maps in panels (c) and (d). Magnetic field lines derived from the same MHD model are overlaid in all panels. It can be seen that the MHD model and the multiwavelength observations yield a qualitative match in the flare geometry and various plasma properties. We note that both the Hinode/EIS and SDO/AIA maps are affected by saturation at the brightest portion of the flare arcade (red dashed contour). Also, the maps show a diffraction pattern (with an ``X'' shape) originating from the brightest region. One should disregard the plasma diagnostics results from these regions corrupted by such effects.

The particle modeling adopts the method described in \citet{Kong2019}. Similar approaches have been performed in several following studies \citep{Kong2020,Kong2022,Kong2022b,Li2022}. We refer interested readers to these works for more detailed descriptions. Briefly, monoenergetic electrons of an initial energy of 20 keV are injected into the MHD model as pseudoparticles with an isotropic angular distribution. The kinetic evolution of these particles in the simulation domain is modeled by solving the Parker transport equation, which takes the fluid velocity and magnetic field input from the MHD model. Particle transport in the magnetic field is mainly modulated by stochastic diffusion by well-developed turbulence with a Kolmogorov-type power spectrum. The construction of the diffusion coefficient follows the treatment in \citet{Giacalone1999}, with the perpendicular diffusion assumed to be 10\% of the parallel diffusion. The simulation domain has an area of $102.0\ \text{Mm}\times 127.5\ \text{Mm}$ with a uniform grid size of 0.22 Mm. The output of the model, binned to a grid size of 1.275 Mm, is a spatial distribution of nonthermal electrons at different energies. In the model, the electron momentum is distributed evenly in logarithmic space, with a total of 40 samples between 20 and 5450 keV.

Figures~\ref{fig:model}(i)--(l) show the distribution of nonthermal electrons at four selected energies (50, 100, 200, and 500 keV) from the particle model. Similar to the results in \citet{Kong2019}, the nonthermal electrons at all energies are strongly concentrated in the magnetic bottle region. As discussed in the main text, we can draw an electron distribution function at each pixel of the model. Figure~\ref{fig:obs_model}(c) shows examples from four selected locations. All the electron spectra display a power-law shape in the $\sim$30--600 keV range, with those in the magnetic bottle region featuring a downward spectral break. Such broken power-law electron distributions in the model are formed by the combination of stochastic acceleration, trapping, and escaping processes \citep{Kong2019,Li2022}. The presence of a downward-breaking electron spectrum in the ALT magnetic bottle region is further supported by recent observations of the same flare during its early impulsive phase using combined EUV, X-ray, and microwave data \citep{Chen2021}. 

\section*{Appendix D: Emission Modeling} \label{append sec:emission}
\begin{figure}[!ht]
\begin{center}
\includegraphics[width=0.5\textwidth]{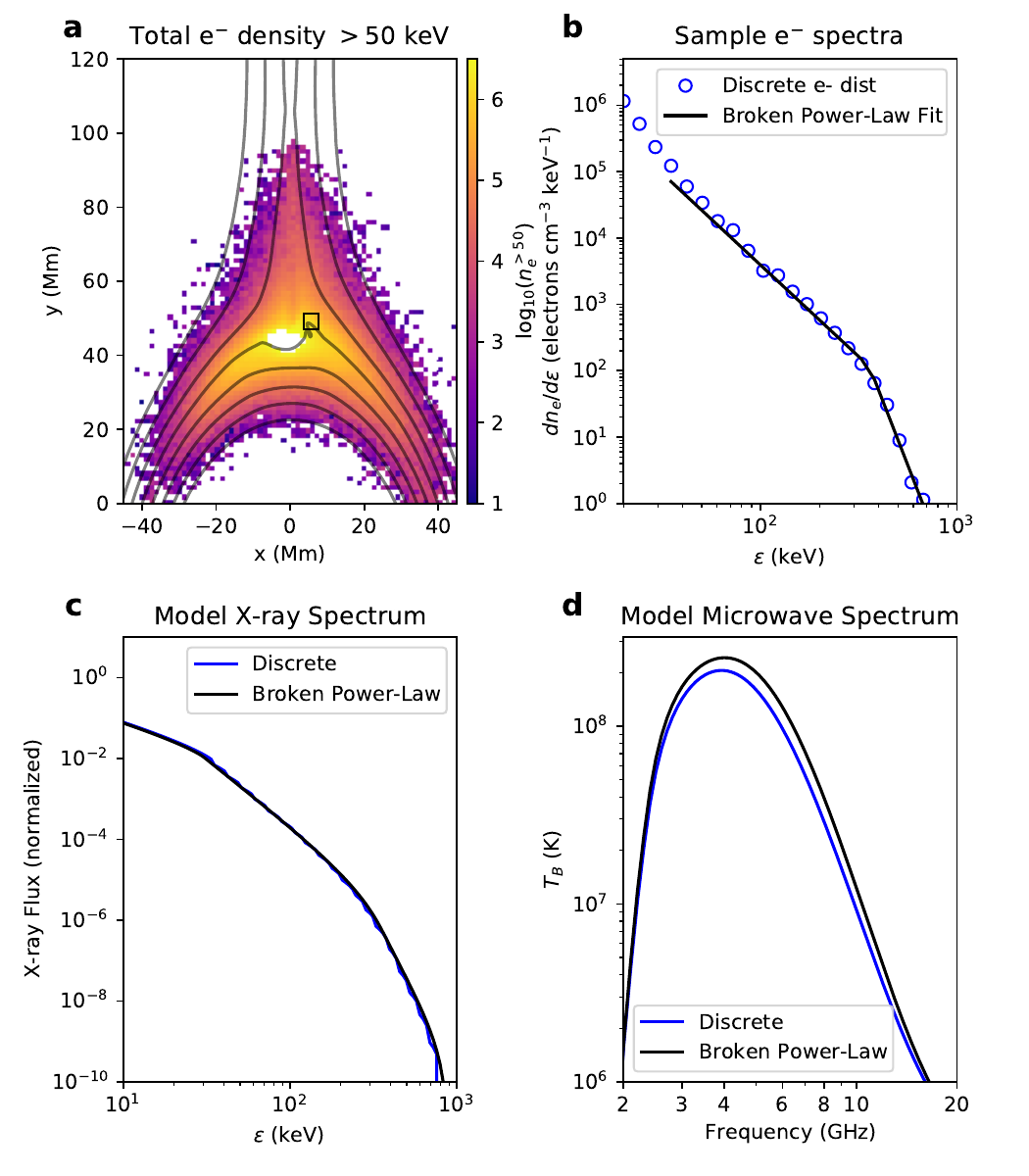}
\end{center}
\caption{Modeled energetic electron distribution and calculated HXR and microwave spectra. (a) Distribution of $>$50 keV energetic electrons from the combined MHD and particle model. (b) Electron distribution from a sample pixel in the model (black box in (a)). Blue circles are the distribution from the model, and the black curve shows the broken power-law fit. (c) Calculated X-ray spectra based on the bremsstrahlung mechanism. Blue and black curves are those using direct integration of the discrete electron distribution and the broken power-law fit, respectively. (d) Similar to (c) but for the calculated microwave spectra based on gyrosynchrotron radiation. }\label{fig:particle_emis}
\end{figure}

With maps of the thermal plasma properties and nonthermal electron distributions, one can calculate synthetic emission maps at various wavelengths and compare them with actual observations. To produce the synthetic SDO/AIA 131 \AA\ EUV map, we only need the plasma density $n_{\rm th}$ and temperature $T$ from the 2.5D MHD model. We first calculate the EUV intensity at each pixel with $I(x, y)=n_{\rm th}^2(x,y)G[T(x,y)]L_z$, where $G(T)$ is the temperature response function of the 131 \AA\ band \citep{ODwyer2010} and $L_z$ is the LOS depth (fixed to be 10 Mm where flare-heated $>8$~MK plasma is present and 1 Mm elsewhere to suppress the coronal background with an artificially high density, as it is not optimized in the model), and then convolve the resulting image using a Gaussian kernel with an FWHM of $1''.2$  according to AIA's point spread function (PSF). Although we do not have sufficient information on the third dimension (along the LOS) and hence have to use a homogeneous assumption, the resulting EUV 131 \AA\ map, shown in Figure~\ref{fig:obs_model}(e), resembles the observations reasonably well (Figure~\ref{fig:obs_model}(f)). 

In order to produce synthetic HXR and microwave maps, both the thermal plasma properties and nonthermal electron distributions are needed. For calculating synthetic HXR maps, at each pixel, we take the plasma density in the MHD model $n_{\rm th}(x, y)$ and the nonthermal electron distribution $dn_e(x,y,\varepsilon)/d\varepsilon$ and compute the X-ray photon flux $I_X(x,y, \epsilon)$ as a function of X-ray photon energy $\epsilon$ based on the thin-target bremsstrahlung theory. To perform the numerical calculations, we have adopted two commensurable approaches. One approach is to first fit the discrete nonthermal electron distribution from the particle model with a broken power-law form, and then supply the best-fit broken power-law form as the input to calculate the expected thin-target bremsstrahlung spectrum using existing tools available from the \texttt{xray} package within the \texttt{sswIDL} distribution\footnote{Description of the software codes can be found at \url{https://hesperia.gsfc.nasa.gov/ssw/packages/xray/doc/brm_thin_doc.pdf}}. Another approach is, for every given X-ray photon energy $\epsilon$, we take the discrete model electron distribution and integrate the X-ray flux contributed by all energy bins numerically. Figure~\ref{fig:particle_emis}(b) shows an example electron distribution derived from a selected pixel in the magnetic bottle region. Blue symbols are the discrete distribution from the particle model, and the black solid line is the best-fit broken power-law function. The calculated X-ray photon spectra using direct integration and a broken power-law fit of the electron distribution are shown in panel (c) as blue and black curves, respectively. The results show that they are in agreement with each other. 

For calculating the synthetic microwave spectrum from each pixel, we use the numerical codes developed by \citet{Kuznetsov2021}, which allow an input electron distribution in both the discrete numerical form and an analytical broken power-law form. Likewise, the results from the two different approaches are generally consistent with each other, although we found that the broken power-law approach gives better-defined microwave spectra as it effectively ``smooths out'' the occasional noise in the input electron distribution, especially in regions with low counting statistics. Therefore, we have adapted the broken power-law fit method to calculate the HXR and microwave spectra pixel by pixel, forming the spectrally resolved HXR and microwave maps. Finally, to compare with the observations, each image is convolved with a Gaussian function with the same FWHM width as the point spread function used to reconstruct the observed RHESSI and EOVSA images ($6''.8$ for RHESSI and $10''.2/[\nu/ {10\ \rm GHz}]$ for EOVSA).

\bibliography{sep10}

\end{document}